\documentclass[reprint,amsmath,amssymb,aps,floatfix,superscriptaddress]{revtex4-1}
\usepackage{xcolor}
\usepackage{physics}
\usepackage{amsmath}
\usepackage{natbib}
\usepackage{fancyhdr} 
\usepackage{lastpage} 
\usepackage{extramarks} 
\usepackage{graphicx} 
    \graphicspath{{Figures/}}
\usepackage{listings} 
\usepackage{courier} 
\usepackage{lipsum} 
\usepackage{float}
\usepackage{resizegather}
\usepackage[titletoc,toc,title]{appendix} 
\usepackage{gensymb}
\usepackage[caption=false]{subfig}
\usepackage{wrapfig}
\usepackage{setspace}
\usepackage{import}
\usepackage{enumerate}
\usepackage{hyperref}
\usepackage{chemformula}
\usepackage{bm}
\usepackage{svg}
\usepackage{dsfont}
\usepackage{enumitem} 
\usepackage[english]{babel}
\usepackage{amssymb}
\usepackage{multirow} 
\usepackage{subfiles} 
\begin{document}
\preprint{APS/123-QED}
\title{Microscopic model of spin flip-flop processes in rare-earth-ion-doped crystals} 

\author{Hafsa Syed}
\affiliation{Department of Physics, Lund University, Lund, Sweden}
\author{Adam Kinos}
\affiliation{Department of Physics, Lund University, Lund, Sweden}
\author{Chunyan Shi}
\affiliation{Department of Physics, Lund University, Lund, Sweden}
\affiliation{Present address: Zurich Instruments, Technoparkstrasse 1, Zurich, Switzerland}
\author{Lars Rippe}
\affiliation{Department of Physics, Lund University, Lund, Sweden}
\author{Stefan Kr{\"o}ll}
\affiliation{Department of Physics, Lund University, Lund, Sweden}

             
\begin{abstract}
 Flip-flop processes due to magnetic dipole-dipole interaction between neighbouring ions in rare-earth-ion-doped crystals is one of the  mechanisms of relaxation between hyperfine levels. Modeling of this mechanism has so far been macroscopic, characterized by an average rate describing the relaxation of all ions. Here however, we present a microscopic model of flip-flop interactions between individual nuclear spins of dopant ions. Every ion is situated in a unique local environment in the crystal, where each ion has different distances and a unique orientation relative to its nearest neighbors, as determined by the lattice structure. Thus, each ion has a unique flip-flop rate and the collective relaxation dynamics of all ions in a bulk crystal is a sum of many exponential decays, giving rise to a distribution of rates rather than a single average decay rate. We employ this model to calculate flip-flop rates in  Pr\(^{3+}\):Y\(_2\)SiO\(_5\) and show experimental measurements of population decay of the ground state hyperfine levels  at $\sim$2 K. We also present a new method to measure rates of individual transitions from hole burning spectra that requires significantly fewer fitting parameters in theoretical rate equations compared to earlier work. Furthermore, we measure the effect of external magnetic field on the flip-flop rates and observe that the rates slow down by two orders of magnitude in a field of 5-10 mT.
\end{abstract}

\maketitle


\section{Introduction} \label{sec:SpinRelaxation}
Rare-earth-ion-doped crystals have hyperfine transitions with unique properties such as long lifetimes and coherence times, for example up to twenty days \cite{konzetal,PhysRevA.98.062516} and six hours  respectively in Eu\(^{3+}\):Y\(_2\)SiO\(_5\) \cite{zhong_optically_2015}. These transitions are easily accessible via optical transitions that are inhomogeneously broadened (up to 100's of GHz) and also possess narrow homogeneous linewidths (\(\leq\) kHz), enabling their use in  quantum memories \cite{nilsson_solid_2005,nicolle_gigahertz-bandwidth_2021,afzelius_impedance-matched_2010} and quantum computing \cite{OHLSSON200271,PhysRevA.101.012309,PRXQuantum.2.010312,PhysRevA.105.032603,Roadmap2021}. Long-lived, optically deep and spectrally narrow holes can be burnt in these materials and they can be used in laser stabilization \cite{sellin_programmable_1999,bottger_programmable_2003,thorpe_frequency_2011,horvath_slow_2022} and as efficient spectral filters in a medical imaging technique called Ultra-sound Optical Tomography \cite{li_pulsed_2008,zhang_slow_2012,xiao_xu_spectral_2010,venet_ultrasound-modulated_2018,hill_acousto-optic_2021}.

Even though hyperfine lifetimes can be as long as seconds or much more, relaxation can be a problem in many of the above applications. It results in decreased absorption depth, leading to lower efficiency of echoes in quantum memories, degrading of spectral filters and decreased gate fidelity in quantum computing. In general, hyperfine relaxation can occur either via lattice vibrations mediated by phonons (Spin-Lattice Relaxation) or via interactions with neighbouring spins (Spin-Spin Relaxation). Spin-Lattice Relaxation processes are well understood \cite{larson_spin-lattice_1966,abragam_electron_2012} and the mechanism relevant at cryogenic temperatures is the Direct process, whose rate increases proportional to the temperature (\(\propto T\)) and square of magnetic field (\(\propto B^2\)) \cite{orbach_spin-lattice_1961}. However, various experiments at cryogenic temperatures have demonstrated a decrease in relaxation rates with the application of a magnetic field \cite{PhysRevB.38.11061,ohlsson_experimental_2003,car_optical_2019,cruzeiro_spectral_2017}. The mechanisms responsible for relaxation in such cases are not phonon related but magnetic dipole interactions between dopants. They are known as flip-flop interactions whereby two nearby ions exchange their spins via magnetic dipole-dipole interaction and the interaction strength wanes with distance \(r\) as \(r^{-6}\). Studies in Kramers ions like Er\(^{3+}\) and Nd\(^{3+}\) have used a macroscopic model to explain spectral hole decay due to flip-flop process by taking a single average rate to be related to the dopant concentration \cite{cruzeiro_spectral_2017} and an average ion-ion distance for the ions in the crystal \cite{car_optical_2019}, resulting in a rate \(R \propto \frac{n^2}{\langle r \rangle ^6}\). However, it has also been reported that this mechanism can lead to non-exponential decays \cite{holliday_spectral_1993} and the focus of this work is to develop a model that captures this effect. 

Each dopant ion in the crystal is randomly placed in the crystalline structure such that it experiences a different magnetic environment and has different distances to and orientations of its nearest neighbouring dopants. Therefore, each ion relaxes with a unique rate and when the relaxation dynamics is studied in a bulk crystal, we see a sum of many exponential decays. In this work, we use a numerical simulation of a host crystal to create a distribution of ion-ion distances. The flip-flop rate between all the pairs of ions is then calculated using Fermi's Golden rule. The shape of this distribution of flip-flop rates mimics that of \(r^{-6}\), where \(r\) is the ion-ion distance. We compare the model to experimental measurements of population decay of hyperfine levels in Pr\(^{3+}\): Y\(_2\)SiO\(_5\). The experiments are done using an alternative method to measure rates of individual transitions using hole-burning spectra. An earlier work \cite{klieber_all-optical_2003} used hole-burning spectra in Pr\(^{3+}\):YAlO\(_3\) to fit 21 parameters to theoretical rate equations. We reduce the number of fitting parameters to 3 by initializing the ground state population in one of the hyperfine levels in a narrow spectral region. Three additional parameters are used to describe the effect of small magnetic fields between 5-10 mT on the flip-flop rates on each of the transitions.

The paper is structured as follows: We first introduce the relaxation pathways for flip-flop interactions considered and enumerate the steps in simulating a distribution of flip-flop rates in Section \ref{sec:dipole_dipole}. We then explain the new experimental method used to measure population decay in Pr\(^{3+}\): Y\(_2\)SiO\(_5\) in Section \ref{sec:experiments}. In Section \ref{sec:results_and_discussion}, we compare the experiments with simulations to extract the flip-flop rates and also show that the distribution of rates arises from a distribution of ion-ion distance. Lastly, we conclude with some comments on further additions to the microscopic model.


\section{Microscopic Model for flip-flop interaction} \label{sec:dipole_dipole}
In this section, we first explain the relaxation pathways and different strengths of each pathway considered in the model, with the example of Pr\(^{3+}\):Y\(_2\)SiO\(_5\). We set up the magnetic dipole-dipole interaction Hamiltonian for a pair of ions and explain the use of Fermi's rule to calculate the flip-flop rate between them. Then, we enumerate the steps in simulating flip-flop interactions and specify the parameters used for  Pr\(^{3+}\):Y\(_2\)SiO\(_5\). 

\begin{figure}[ht]
\centering
\subfloat[]{\includegraphics[width=0.9\linewidth]{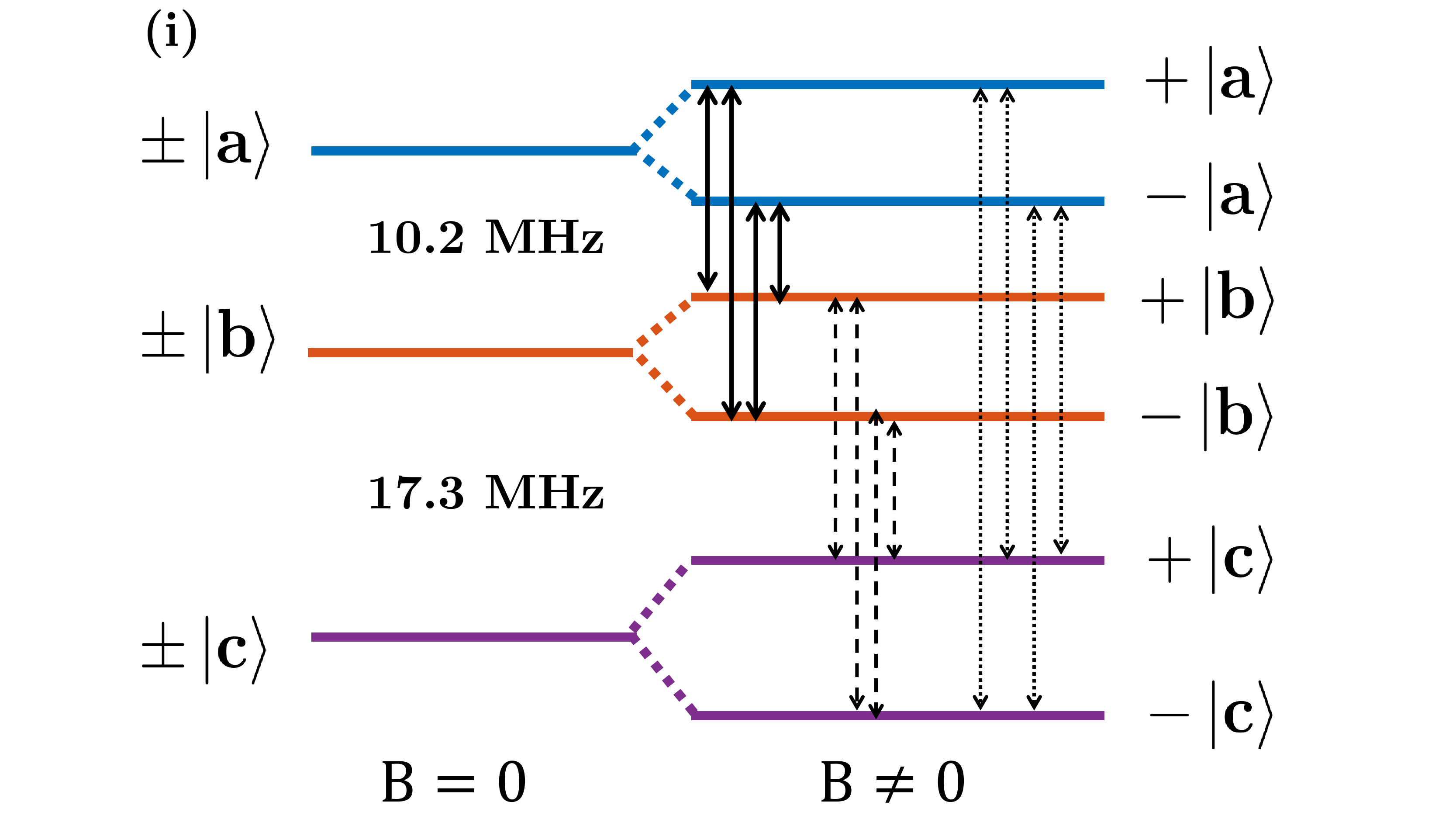}}
  \hfill
  \vfill
  \subfloat[]{\includegraphics[width=1\linewidth]{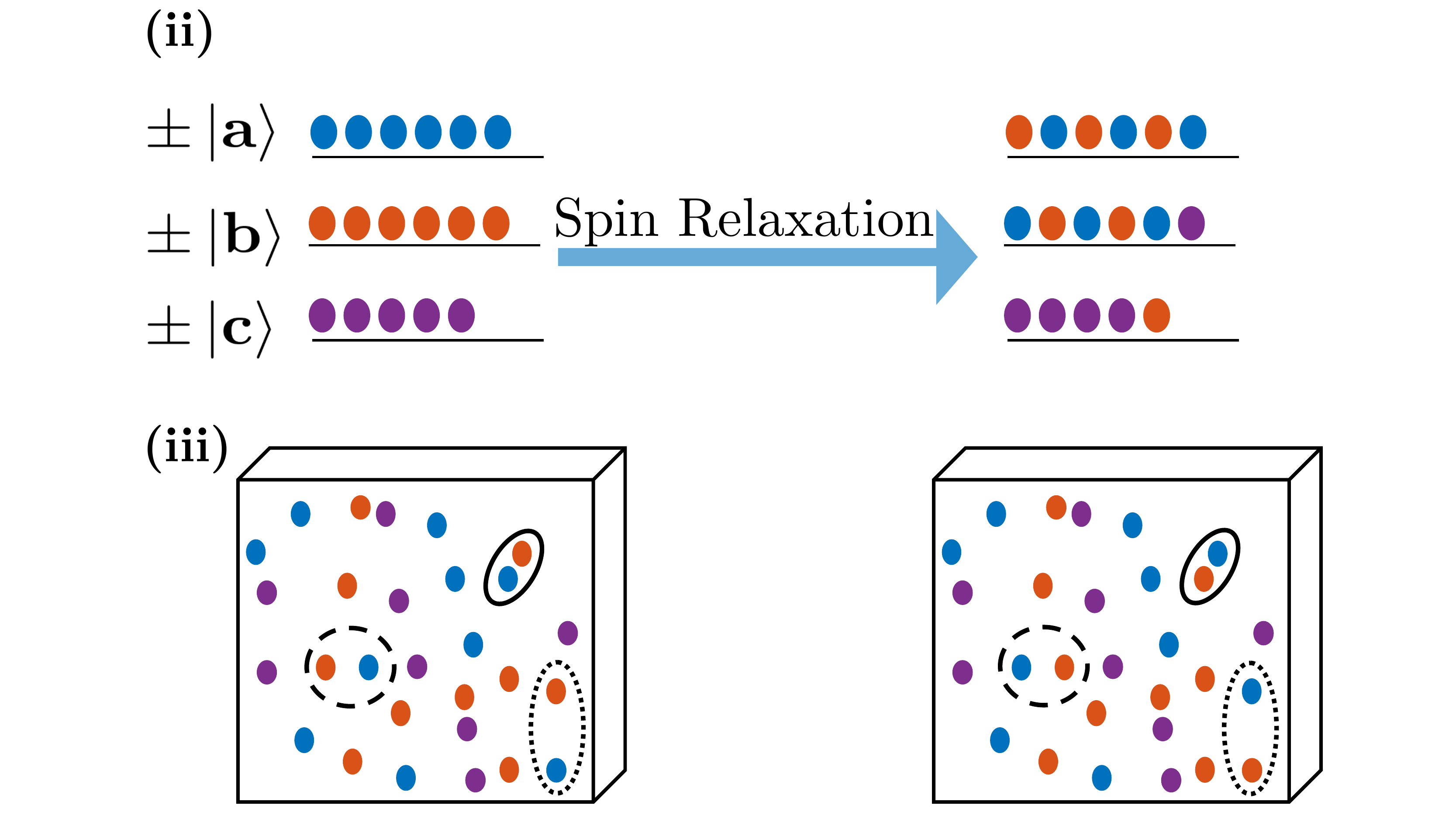}}
 \caption{Representation of spin relaxation via flip-flop interactions in Pr\(^{3+}\):Y\(_2\)SiO\(_5\), whereby two neighbouring ions interchange their state. (i) Ground state hyperfine levels \(\pm \ket{a}\), \(\pm \ket{b}\), \(\pm \ket{c}\) lose their degeneracy in the presence of an external magnetic field B, forming six levels. The pathways considered for spin relaxation in our simulations are shown with double-sided arrows. (ii) Ions occupying \(\pm \ket{a}\) and \(\pm \ket{b}\) flip-flop strongly with each other while those in \(\pm \ket{c}\) share weak interaction with either of the other hyperfine levels. (iii) The interaction strength varies with distance \(r\) as \(r^{-6}\) so closely-lying neighbours in the crystal can rapidly flip-flop (shown as solid ovals) while ions separated by larger distances share weaker interaction (shown as dashed and dotted ovals).}
  \label{fig:spinspin interaction}
\end{figure}

Figure \ref{fig:spinspin interaction}(i) shows the three hyperfine levels in the electronic ground state ${}^{3}_{}H^{}_{4}$ Pr\(^{3+}\):Y\(_2\)SiO\(_5\) and the twelve relaxation pathways considered in the simulations. Each hyperfine level is doubly degenerate but the degeneracy is lost in the presence of an external magnetic field, giving rise to six levels in total. It should be noted that conventional labels for the hyperfine levels are \(\pm\ket{\frac{1}{2}g}, \pm\ket{\frac{3}{2}g}, \pm\ket{\frac{5}{2}g}\) but each level is in reality an admixture of all six hyperfine wave functions. So we instead use the labels \(\pm\ket{a}, \pm\ket{b}, \pm\ket{c}\).  For six levels, we could expect fifteen unique flip-flop transitions. But we do not consider the transitions where only the parity changes, for example transitions of the type \(+\ket{a} \leftrightarrow -\ket{a}\) since we do not measure these individually in our experiments. Hence, we have twelve pathways in total. In the experiments described in Section \ref{sec:experiments}, it is seen that the strongest interaction is between ions occupying \(\pm \ket{a}\) and \(\pm \ket{b}\) (indicated by solid double-sided arrows), while \(\pm \ket{c}\) couples weakly to the other two levels (indicated by dashed and dotted double-sided arrows). Figure \ref{fig:spinspin interaction}(ii) shows the dependence of strength of flip-flop interaction on the hyperfine level occupied by ions. For example, ions initialized in \(\pm \ket{a}\) and \(\pm \ket{b}\) flip-flop strongly to give fairly mixed populations (blue and red circles) while ions in \(\pm \ket{c}\) (purple circles) flip-flop with either of the other two levels with less likelihood. Figure \ref{fig:spinspin interaction}(iii) visualizes how the interaction strength scales with distance as \(r^{-6}\), thus closely-lying neighbours in a crystal interact strongly (shown as solid ovals) while ions far away from each other show weaker interaction (shown as dashed and dotted ovals).

The rate for ion `\(i\)' to flip from \(\ket{x}\) to \(\ket{y}\) due to interactions with its `\(j\)' neighbours initially in the state \(\ket{y}\) is calculated using Fermi's Golden Rule :
\begin{align} 
\begin{split} \label{Fermi rule}
     R^{i}_{\ket{x} \rightarrow \ket{y}} = & \frac{2\pi}{\hbar} \sum_{j}\\ & |\matrixelement{y^i\otimes x^j}{H_{dd}^{ij}}{x^i\otimes y^j}|^2  f(E)  \\
\end{split}
 \end{align}

It is worth noting that calculation of the matrix elements \(|\matrixelement{y^i\otimes x^j}{H_{dd}^{ij}}{x^i\otimes y^j}|\) for all pair of ions makes our model `microscopic', setting it apart from previous similar works, where this was taken as an average value and related to the concentration of dopants in Ref.\cite{car_optical_2019,cruzeiro_spectral_2017}. 
 
 Wavefunctions of hyperfine levels \( \ket{x^i}\) and \(\ket{y^j} \) are the eigenstates of the spin Hamiltonian and they depend on the external magnetic field \(\mathbf{B}\). They are calculated using the following equation for the spin Hamiltonian, as used in \cite{lovric_spin_2012} :
\begin{equation} \label{spin hamiltonian}
    H_{spin} = \mathbf{B}.\mathbf{M}.\Tilde{\mathbf{I}} + \Tilde{\mathbf{I}}.\mathbf{Q}.\Tilde{\mathbf{I}}
\end{equation}

 The crystallographic axes of the crystal \(\mqty[D_1 & D_2 & b]\) form the common frame of reference for the above calculations. \(\Tilde{\mathbf{I}}\) is the vector of nuclear spin operators \(\Tilde{I_x}, \Tilde{I_y}, \Tilde{I_z}\) and \(\mathbf{B}\) is the magnetic field vector. \(\mathbf{M}\) is the effective Zeeman tensor and  \(\mathbf{Q}\) is the effective quadrupole tensor, defined as follows :
\begin{equation} \label{M matrix}
\mathbf{M} = \mathbf{R_M} . 
\begin{bmatrix}
g_x & 0 & 0 \\
0 & g_y & 0\\
0 & 0 & g_z
\end{bmatrix} 
.\mathbf{R_M}^T = \begin{bmatrix}
g_{xx} & g_{xy} & g_{xz} \\
g_{yx} & g_{yy} & g_{yz}\\
g_{zx} & g_{zy} & g_{zz}
\end{bmatrix} ,
\end{equation}

\begin{equation} \label{Q matrix}
\mathbf{Q} = \mathbf{R_Q}. 
\begin{bmatrix}
E- \frac{1}{3}D & 0 & 0 \\
0 & -E- \frac{1}{3}D & 0\\
0 & 0 & \frac{2}{3}D
\end{bmatrix} 
.\mathbf{R_Q}^T,
\end{equation}
Each of the above matrices is transformed into the frame \(\mqty[D_1 & D_2 & b]\) using rotation matrices with appropriate Euler angles : \(\mathbf{R_k} = R(\alpha,\beta,\gamma)\). The two terms on the right-hand side of Equation (\ref{spin hamiltonian}) are evaluated according to \cite{abragam_electron_2012}: \(\mathbf{B}.\mathbf{M}.\Tilde{\mathbf{I}} = g_{pq}B_p\Tilde{I_q}\) and \(\Tilde{\mathbf{I}}.\mathbf{Q}.\Tilde{\mathbf{I}} = Q_{pq}\Tilde{I_p}\Tilde{I_q}\) where \(p,q = x,y,z\) and the usual summation rules are to be observed whenever a suffix occurs twice. 

\(H_{dd}^{ij}\) is the Hamiltonian for magnetic dipole - dipole interaction between an ion `\(i\)' and a neighbouring ion `\(j\)' \cite{abragam_electron_2012}. 
 \begin{align} 
 \begin{split}\label{dipole dipole hamil}
     H_{dd}^{ij} = & \frac{\mu_0 \hbar^2}{4\pi} \Tilde{I}_{p}^i \Tilde{I}_{q}^i \big\{ g_{ps}^ig_{qs}^j - \frac{3\mathbf{r}^{ij}_s \mathbf{r}^{ij}_t}{|\mathbf{r}^{ij}|^2} g_{ps}^ig_{qt}^j\big \} \frac{1}{|\mathbf{r}^{ij}|^3} \\
     \end{split}
 \end{align} 
 
where each of the suffixes \(p,q,s,t\) take the values \(x,y,z\). \(\mathbf{r}^{ij}\) is the vector connecting the two ions. 

The last factor in Fermi's golden rule in Equation (\ref{Fermi rule}) is \(f(E)\), the density of initial and final states for transitions between two levels in the continuum of initial and final states \(\ket{x^i}\otimes\ket{y^j}\) and \(\ket{y^i}\otimes\ket{x^j}\) respectively. The form of density of states we use is \(f(E) = \frac{1}{\pi h} \frac{\Gamma_{hom}(\mathbf{B})}{\Gamma^2_{hom}(\mathbf{B}) + [\kappa_{xy}(\mathbf{B})\Gamma_{xy}]^2}\), where \(\Gamma_{hom}\) and \(\Gamma_{xy}\) are the homogeneous and inhomogeneous linewidths of the transition \(\ket{x} \leftrightarrow \ket{y}\). \(\Gamma_{hom}\) is a function of external magnetic field \(\mathbf{B}\) and  \(\kappa_{xy}(\mathbf{B})\) is a phenomenological addition to describe the increase in inhomogeneous linewidths in the presence of a magnetic field. The details of derivation of density of states is given in Appendix \ref{densityofstates}. In principle, all pairs of ions are spectrally separated by a different value in the distribution of spin inhomogeneous broadening \(\Gamma_{xy}\). However, here we focus on the microscopic effect of distances between the ions being different and take an average value for \(\Gamma_{xy}\).

 In brief, the simulation steps required for calculating flip-flop rates are :
\begin{enumerate}
    \item A small sphere of a host crystal is simulated, where ions are placed according to the crystal lattice structure \cite{villars_y2sio5_nodate}. It is doped with a rare-earth ion with the specified concentration. Alternatively, one could also assume a continuous random distribution function of ions to determine the position of nearest N\(^{th}\) neighbour, as done in \cite{gomes_cross-relaxation_1996}. More details about modelling the host crystal can be found in \cite{PhysRevA.105.032608}. An ion `\(i\)' is picked in the sphere and nearest neighbours `\(j\)' are found. Nuclear wave functions \(+\ket{a^i}..-\ket{c^i}\) and \(+\ket{a^j}..-\ket{c^j}\) are calculated to be eigenstates of the spin Hamiltonian in Equation (\ref{spin hamiltonian}) and depend on the orientation of the ion in the crystal and the magnetic field.
    \item The dipole-dipole interaction Hamiltonian for ion `\(i\)' due to interaction with neighbours `\(j\)' is calculated according to Equation (\ref{dipole dipole hamil}).
    \item Flip-flop rates for the transitions between all hyperfine levels are calculated using Fermi's rule in Equation (\ref{Fermi rule}). 
    \end{enumerate}
    
    In Pr\(^{3+}\): Y\(_2\)SiO\(_5\), only the ions in site 1 corresponding to the ${}^{3}_{}H^{}_{4}$ \(\rightarrow\) ${}^{1}_{}D^{}_{2}$ transition at 606 nm were used. The radius of sphere used was 100 nm and the flip-flop rate of ion `\(i\)' was calculated due to the interaction with its twenty nearest neighbors. Pr\(^{3+}\) has a nuclear spin \(\frac{5}{2}\), thus \(\mathbf{\Tilde{I}}\) is a (3 \(\times\) 1) vector where each element is a (6 \(\times\) 6) matrix. The eight basic molecules in a unit cell of Y\(_2\)SiO\(_5\) have four different directions so for any ion `\(i\)', the tensors \(\mathbf{M}\) and  \(\mathbf{Q}\) in Equation [\ref{spin hamiltonian}] have one of the four orientations. Values for all the parameters in Equation (\ref{M matrix}) and (\ref{Q matrix}) were taken from Raman Heterodyne Spectroscopy measurements done in Ref.\cite{lovric_spin_2012}. The magnetic field was directed along the crystal axis \(b\). Homogeneous linewidths \(\Gamma_{hom} = \frac{1}{\pi T_2}\) were taken Ref. \cite{PhysRevLett.92.077601,FRAVAL2004347} where the spin coherence time \(T_2\) was measured to be 0.5 ms with zero magnetic field and 6 ms in the presence of magnetic field of 2mT. It does not change appreciably even up to 100 mT, so 6 ms was used for the data with a field between 5-10 mT. The values for all the individual transitions have not been measured, so the same was used for all. Furthermore, our experiments do not distinguish between the rates of the form of \(+\ket{a} \rightarrow +\ket{b}\) from \(+\ket{a} \rightarrow -\ket{b}\), \(-\ket{a} \rightarrow +\ket{b}\) or \(-\ket{a} \rightarrow -\ket{b}\), so we in the following sections sum and average the rates such that only three effective rates \(R_{ab},R_{bc}\) and \(R_{ac}\) were obtained for each ion `\(i\)'. The details of reducing twelve rates down to three are described in Appendix \ref{reduction_of_rates}. After this reduction, the model contains six unknowns: the three inhomogeneous spin linewidths \(\Gamma_{ab}, \Gamma_{bc},\Gamma_{ac}\) and the factors describing their magnetic field dependence \(\kappa_{ab}, \kappa_{bc}, \kappa_{ac}\) used in the density of states f(E).

\section{Experiments} \label{sec:experiments}
Relaxation between spin levels has been studied in many different ways, for example using methods that combine optical spectral hole burning and RF fields resonant with a hyperfine transition \cite{shelby_optical_1980,bohan_temperature_1969, blasberg_nuclear_1993}. A method to extract rate constants for individual transitions using only hole burning spectra has been used in \cite{klieber_all-optical_2003} but it requires many fitting parameters for each rate equation to be able to keep track of the initial population of any ion that was excited during the hole burning. For example,  Pr\(^{3+}\):YAlO\(_3\) has three hyperfine levels in the ground and excited states. Thus, a laser at a single frequency on the ${}^{3}_{}$H$^{}_{4}$ \(\rightarrow\) ${}^{1}$D$^{}_{2}$ transition can excite nine different transitions or classes of ions. So the method in Ref.\cite{klieber_all-optical_2003} required 21 independent fitting parameters (18 initial spin populations and 3 rates). Here, we present an alternative method to measure individual transition rates by initializing population in one hyperfine level (or, equivalently in a single class) within a narrow spectral region and tracking the decay of this state-specific hyperfine population versus its neighbouring spectral background. This method can be advantageous for measurements in rare-earth ions with more than one ground hyperfine level, where there are multiple classes of ions since the number of parameters for initial spin population are reduced due to initialization. 

\begin{figure}[ht]
\centering
 \subfloat[]{\includegraphics[width=0.5\linewidth]{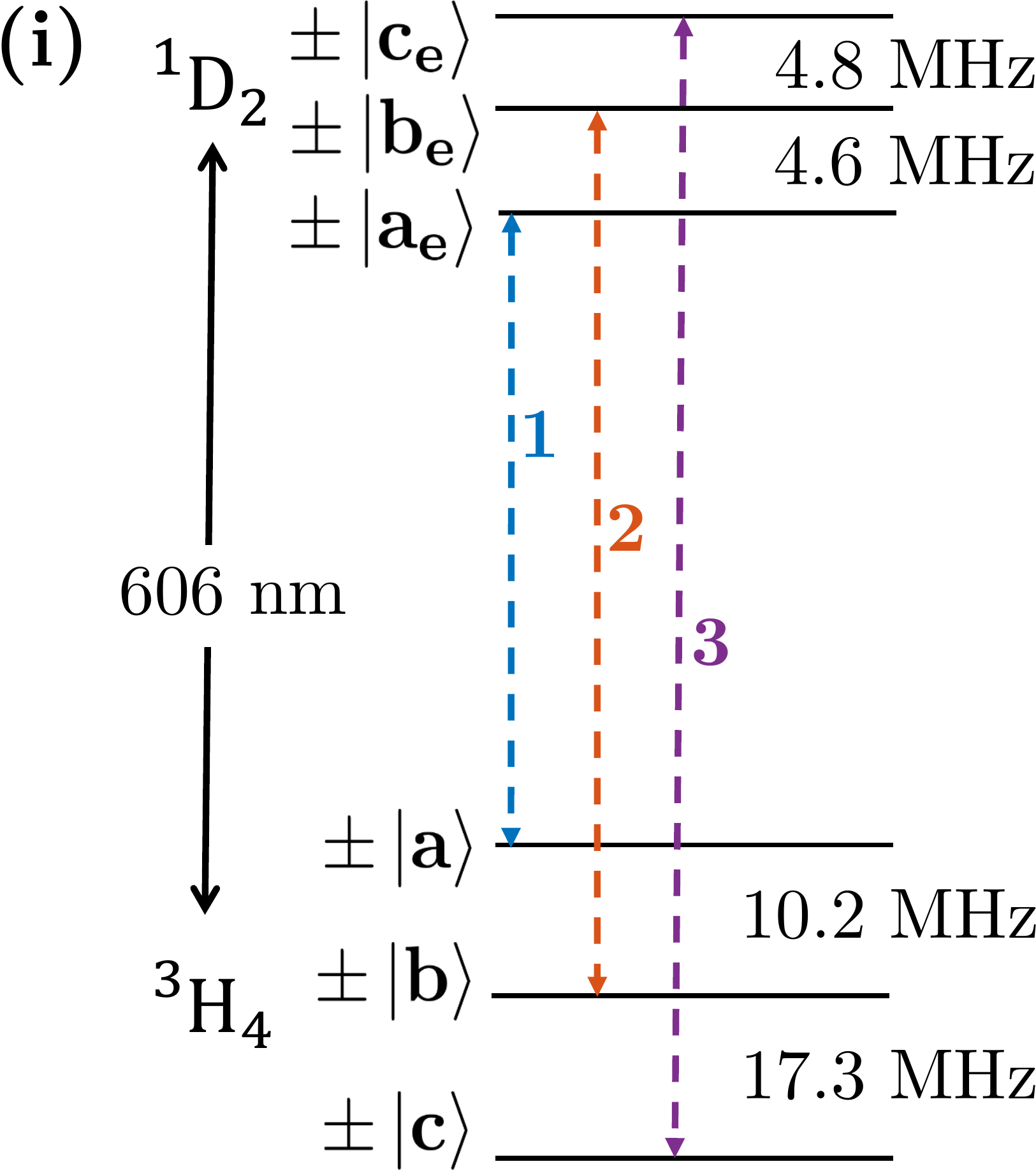}}
  \hfill
  \vfill
\subfloat[]{\includegraphics[width=0.9\linewidth]{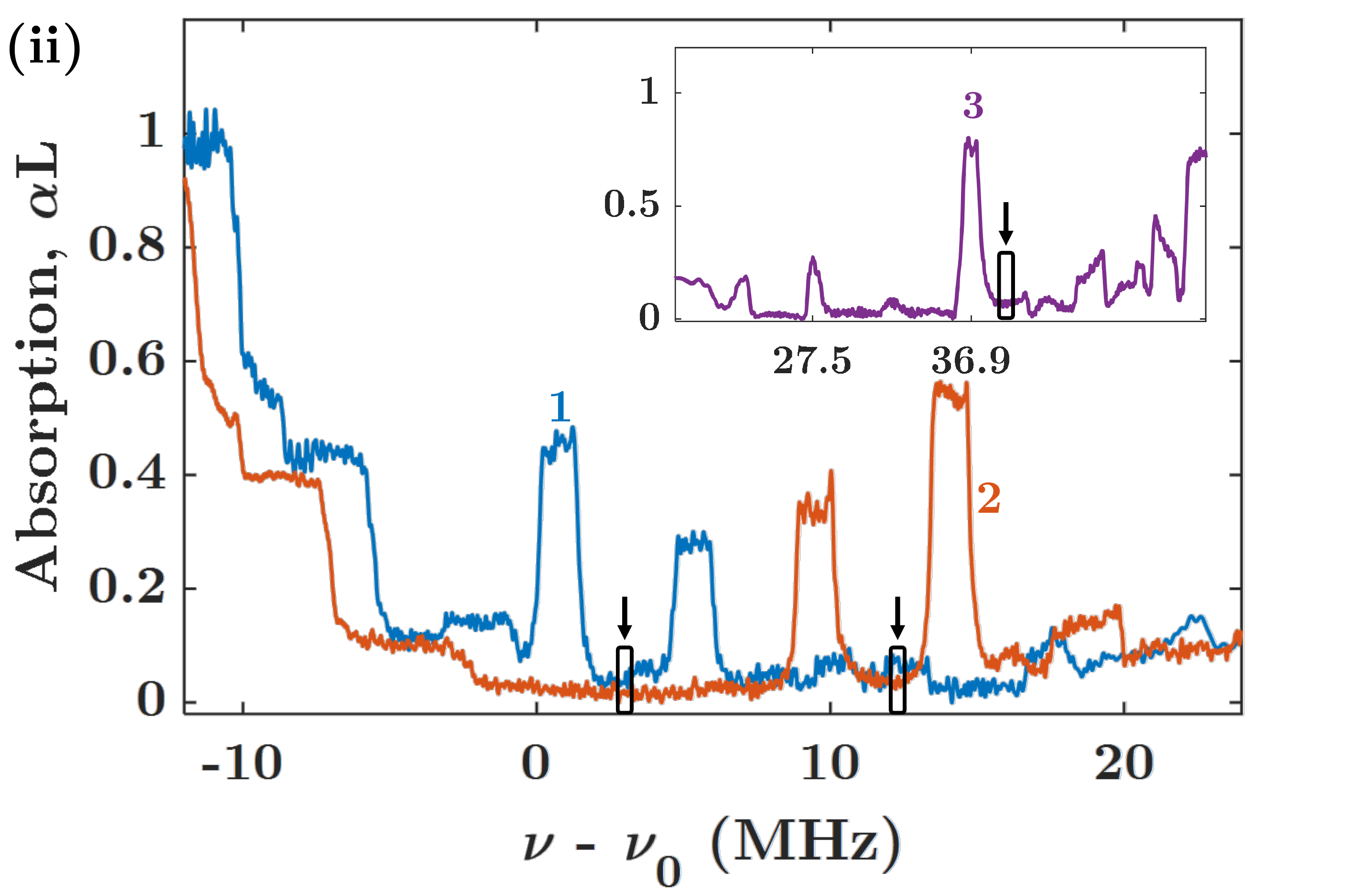}}
 \caption{ Hyperfine energy levels in Pr\(^{3+}\):Y\(_2\)SiO\(_5\) corresponding to the  ${}^{3}_{}H^{}_{4}$ \(\rightarrow\) ${}^{1}_{}D^{}_{2}$ transition at 606 nm and the absorption spectrum obtained from the experiments. (i) Transitions used for evaluating population decay in this experiment are labelled as `1', `2', `3' (ii) Absorption spectrum after initializing the population in \(\pm \ket{a}\) (blue), \(\pm \ket{b}\) (red) and  \(\pm \ket{c}\) (inset, purple). Peaks labelled as `1', `2', `3' correspond to the transition shown in (i). The background absorption region considered for each of the peaks is marked with black arrows at 2 MHz, 12.2 MHz and 38.9 MHz.}
 \label{fig:energylevels}
\end{figure}

We now describe the steps in experiments. We first create a transmission window using spectral tailoring techniques as described in Ref.\cite{nilsson_hole-burning_2004} and initialize the population in one of the ground state hyperfine levels within a spectral region of 1 MHz inside the window. This enables coupling of the laser to a single class of ions and appropriate selection of spectral background range enables us to monitor only this class of ions rather than all the nine classes. Initial population conditions and evolution for all classes of ions are explained in Appendix [\ref{sec:rateeqns}]. By probing the ions at different intervals of time, we recorded decay curves for each of the levels, up to 2700s.  The absorption structure was erased and the population was reset using a strong frequency scanning pulse after the last readout. The transmission window was then recreated. Experiments were also carried out in the presence of an external magnetic field in the range 5-10 mT, along the crystal axis `b'. For each experiment, the field was turned on after the step of population initialization. For a given hyperfine level, the population decays at the same rate (within \(\pm\) 5-10 \%) in the range 5-10 mT. Thus, we take the average of the decay for each hyperfine level for this range of magnetic fields. 

\begin{figure}
         \centering
         \includegraphics[width=0.9\linewidth]{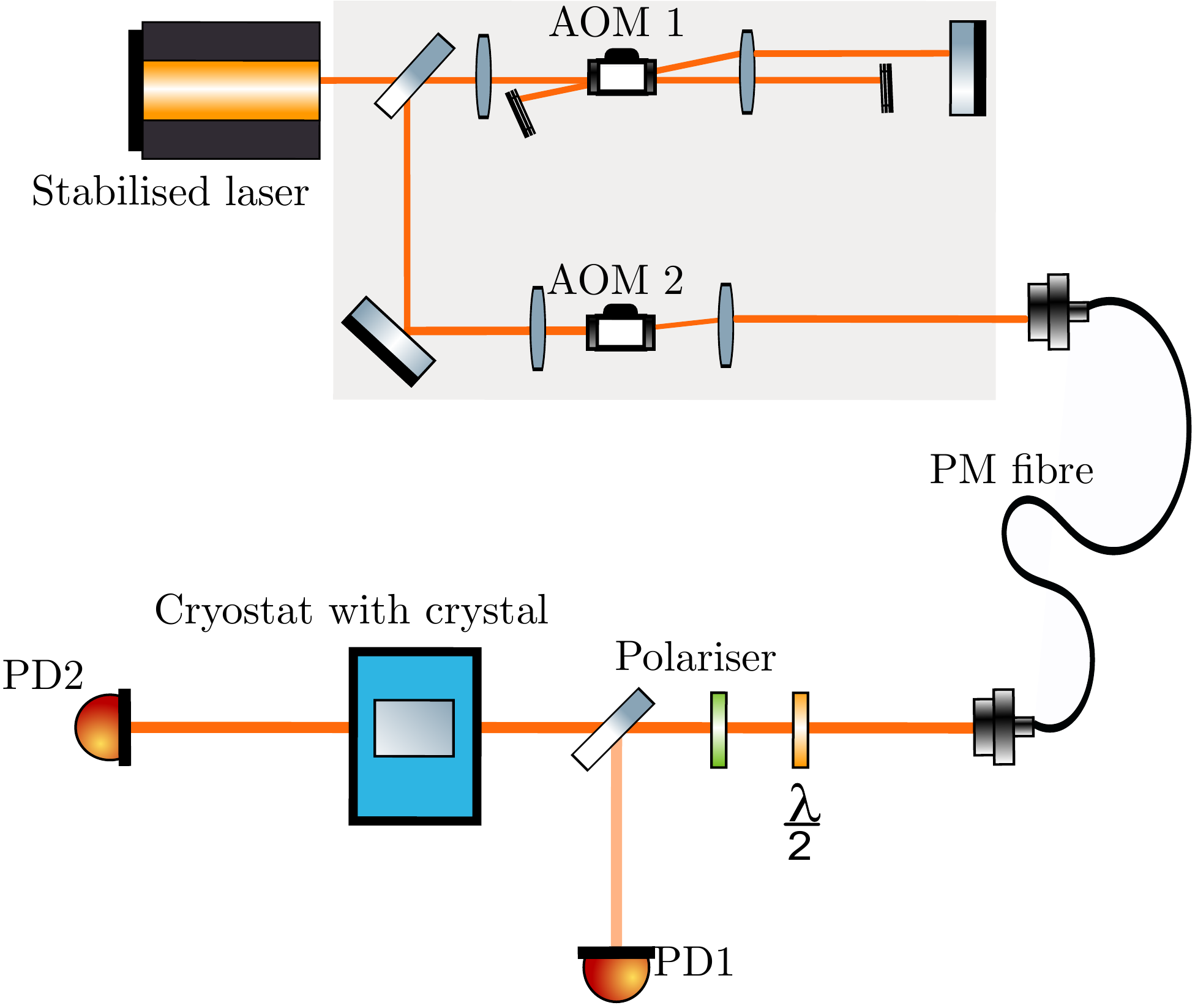}
         \caption{Stabilized laser light is modulated in frequency and amplitude using `AOM 1' (center frequency 200 MHz) in double pass configuration and `AOM 2' (center frequency 60 MHz). A polarization maintaining fibre guides light onto a different table with the cryostat. Polarization of light is adjusted to be along \(D_2\) axis using a polarizer and half-wave plate. A portion of incoming light is sent to a reference detector (PD1) and transmission through the crystal is detected by the transmission detector (PD2).}
         \label{fig:setup}
\end{figure}

Transitions used for evaluating the population and an example of the absorption structure after the initialization process within 1 MHz region are shown in Figure \ref{fig:energylevels}. The optical transitions labelled as `1', `2', `3' in (i) have higher oscillator strength than other transitions, thus the corresponding absorption peaks in (ii) show high absorption and are used for data analysis. Three spectra in blue, red and purple (inset) show the absorption spectrum after initializing ions in \(\pm \ket{a}\),\(\pm \ket{b}\) and \(\pm \ket{c}\) respectively. Evaluation of population is done in two steps. First, a slope is subtracted across the width of each peak since the background on either side might be different on the low and high frequency sides of the peak. This can be seen, for example in peak `1' in Figure \ref{fig:energylevels}(ii). Second, the area under the peaks labelled `1' is summed up to obtain the population \(\pm \ket{a}\) and the same is done for peaks `2' and `3' to obtain populations in \(\pm \ket{b}\) and \(\pm \ket{c}\) respectively. Background absorption level is indicated with black arrows at 2 MHz, 12.2 MHz and 38.9 MHz in Figure \ref{fig:energylevels}. More details can be found in Appendix \ref{sec:spectrawithwithoutB}.

All experiments were done in a Pr\(^{3+}\):Y\(_2\)SiO\(_5\) crystal with 0.05\% concentration and dimensions 10mm x 10mm x 0.8mm along D\(_1\),D\(_2\),b axes respectively. The crystal was placed inside a liquid helium bath cryostat and cooled down to $\sim$2 K. The light source was a dye laser tuned to the ${}^{3}_{}$H$^{}_{4}$ \(\rightarrow\) ${}^{1}$D$^{}_{2}$ transition in Pr\(^{3+}\):Y\(_2\)SiO\(_5\) at 606 nm and was locked to an ultra low expansion glass cavity using the Pound-Drever-Hall locking technique, reducing the linewidth to sub-kHz. A schematic of the experimental setup is shown in Figure \ref{fig:setup}. All the pulses were shaped using an arbitrary waveform generator and two AOMs. A half-wave plate in combination with a polarizer aligns the polarization of the light to the \(D_2\) axis of the crystal with an absorption coefficient measured to be $40$ cm$^{-1}$. The optical power of light for burning pulses was about 20 mW. The readout probe had sufficiently low power such that the same absorption structure could be read up to 100 times without disturbing the population. This was checked by reducing the power until the change in absorption after 100 readouts was within shot-to-shot fluctuations. A collimated 1 mm diameter beam, propagating along the b axis (0.8 mm) of the crystal was used. More details about experiments are described in Appendix \ref{sec:experimental_details}.

 \begin{figure}
         \centering
\subfloat[]{\includegraphics[width=0.8\linewidth]{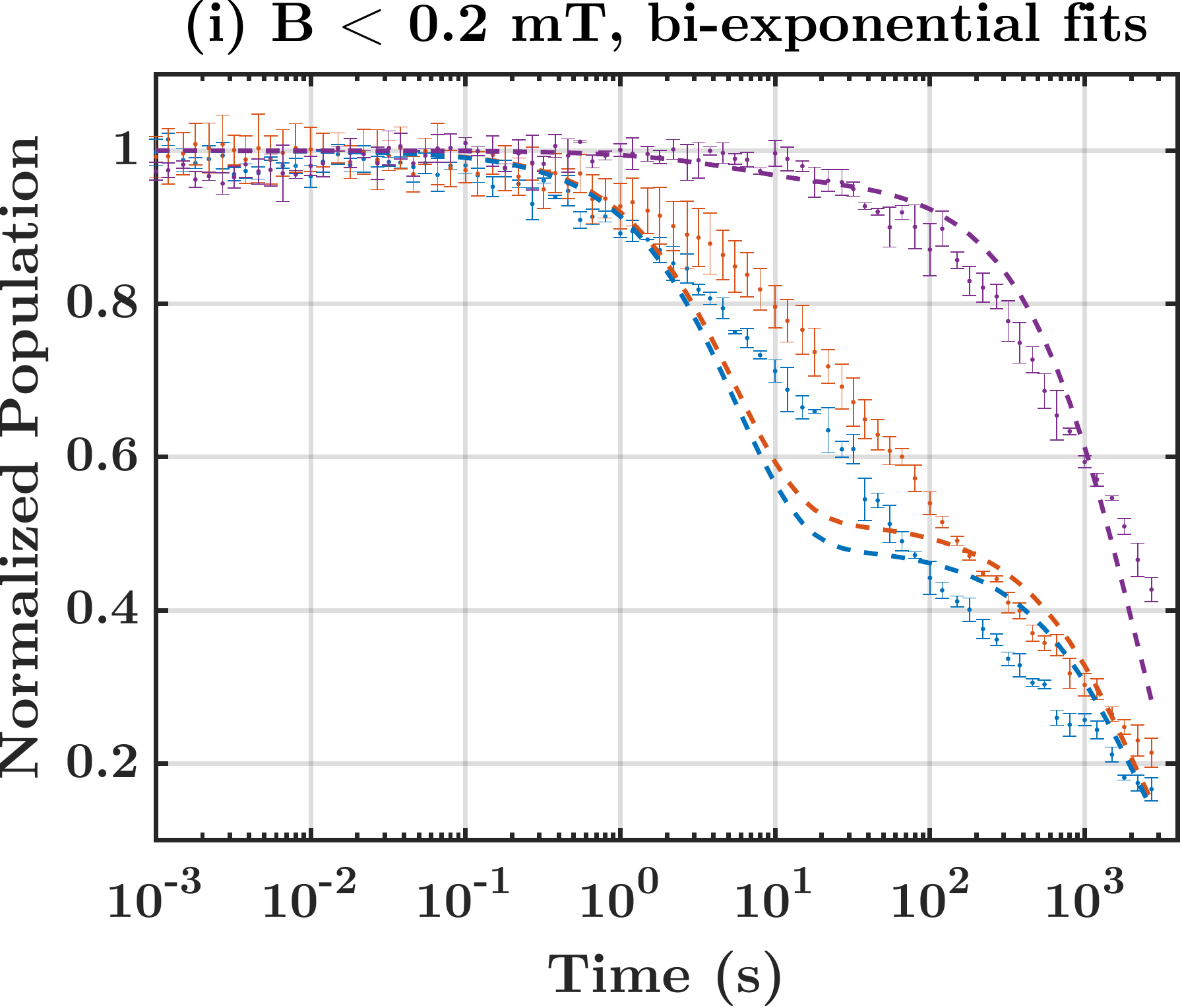}}
          \vfill
     \vfill
     \subfloat[]{\includegraphics[width=0.8\linewidth]{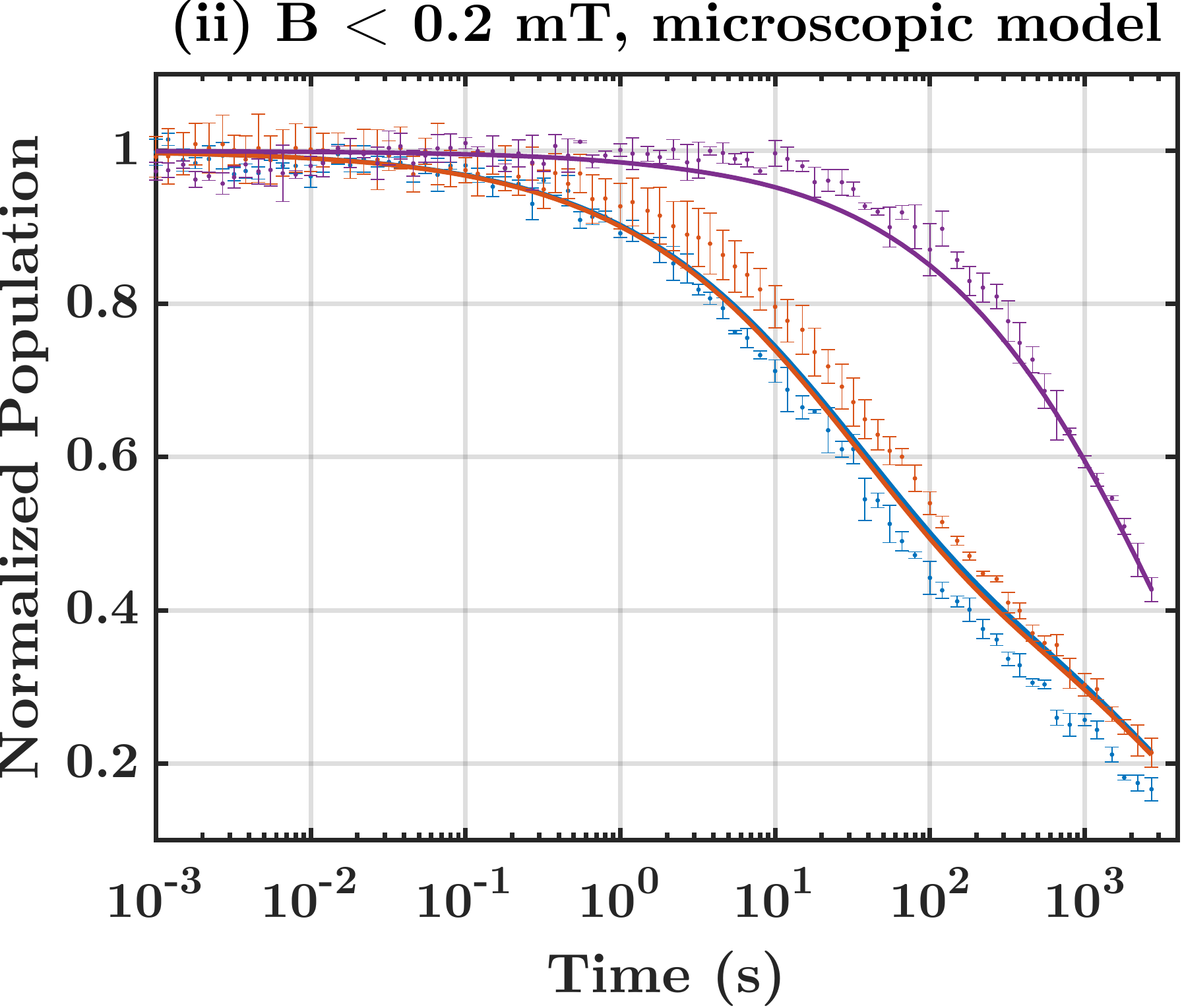}}
         \vfill
     \vfill
     \subfloat[]{\includegraphics[width=0.8\linewidth]{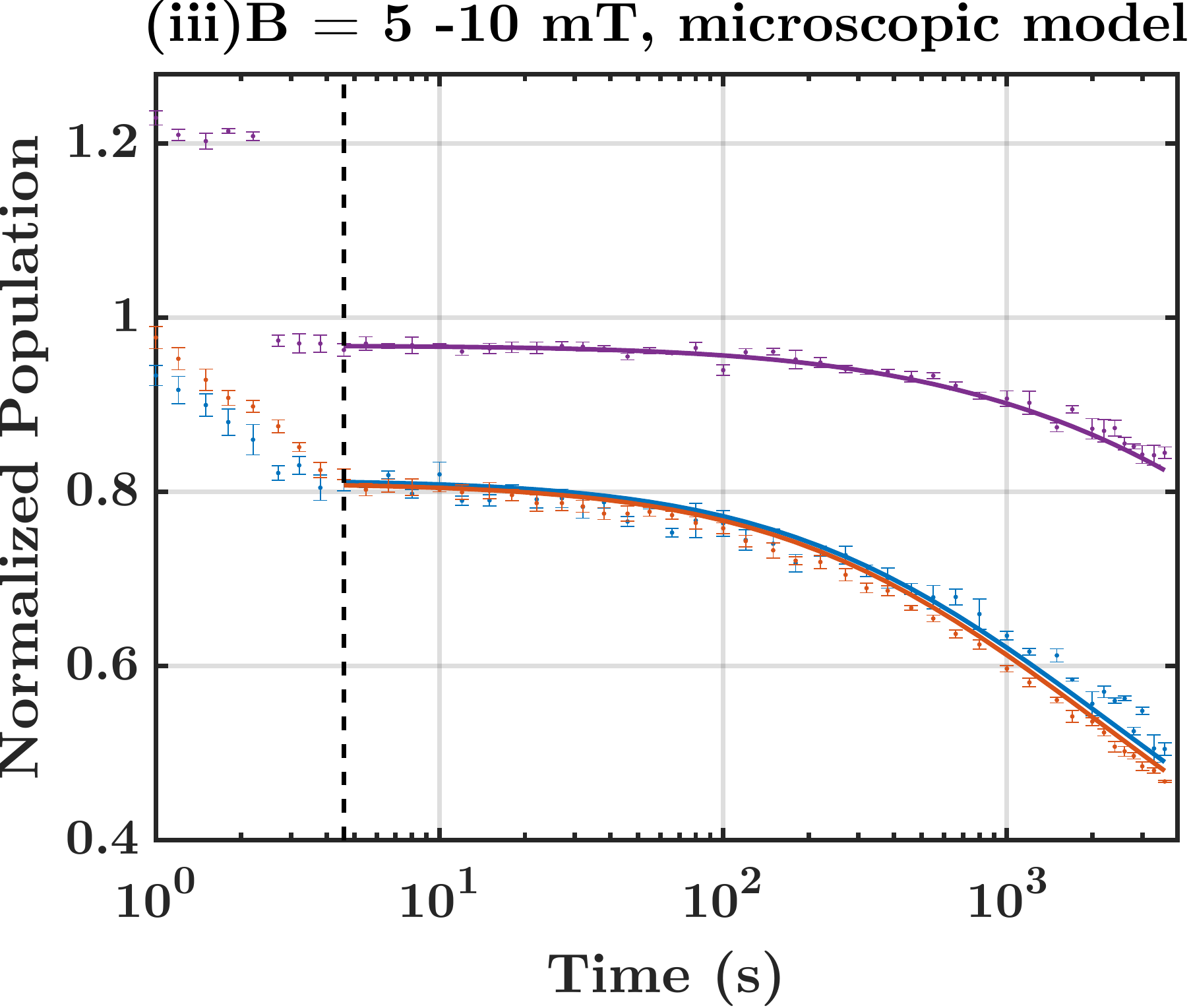}}
          \caption{Population decay of three ground states are shown in three colors : \(\pm \ket{a}\) in blue, \(\pm \ket{b}\) in red and \(\pm \ket{c}\) in purple. The experimental data is shown with errorbars. (i) The colored dashed-dotted lines are the best bi-exponential fits. (ii) The colored solid lines are the result of simulation using our microscopic model described in Section \ref{sec:dipole_dipole}. (iii) Experimental data and microscopic model simulations with external field of 5-10 mT. The vertical dashed black line shows the time \(t_0\) at which \(\vec{B}\) is reached after turning it on. See text for details.}
       \label{fig:DecayFits}
               \end{figure}

\begin{figure*}
\centering
\subfloat[]{\includegraphics[width=0.4\linewidth]{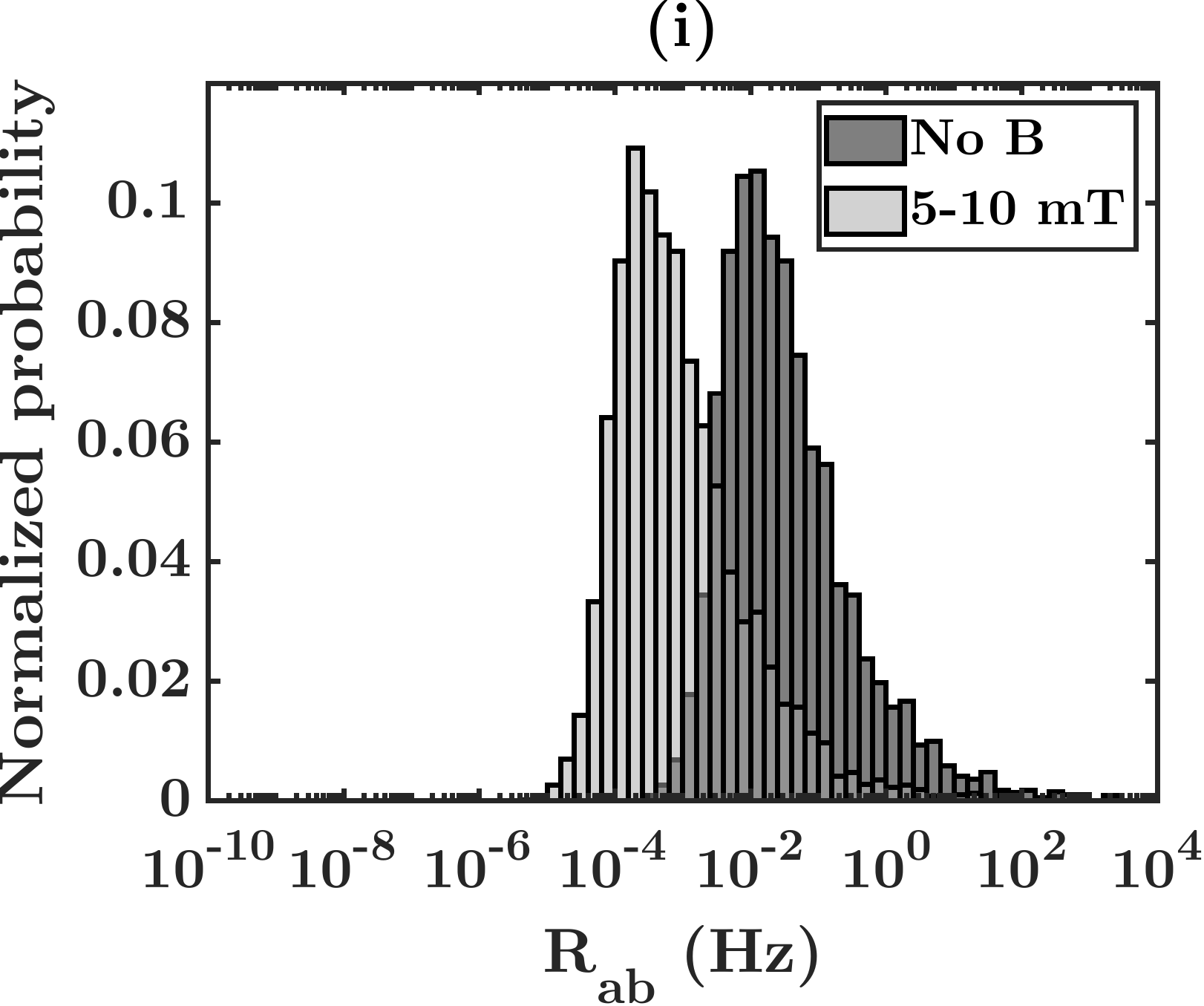}}
\hspace{1cm}
\subfloat[]{\includegraphics[width=0.4\linewidth]{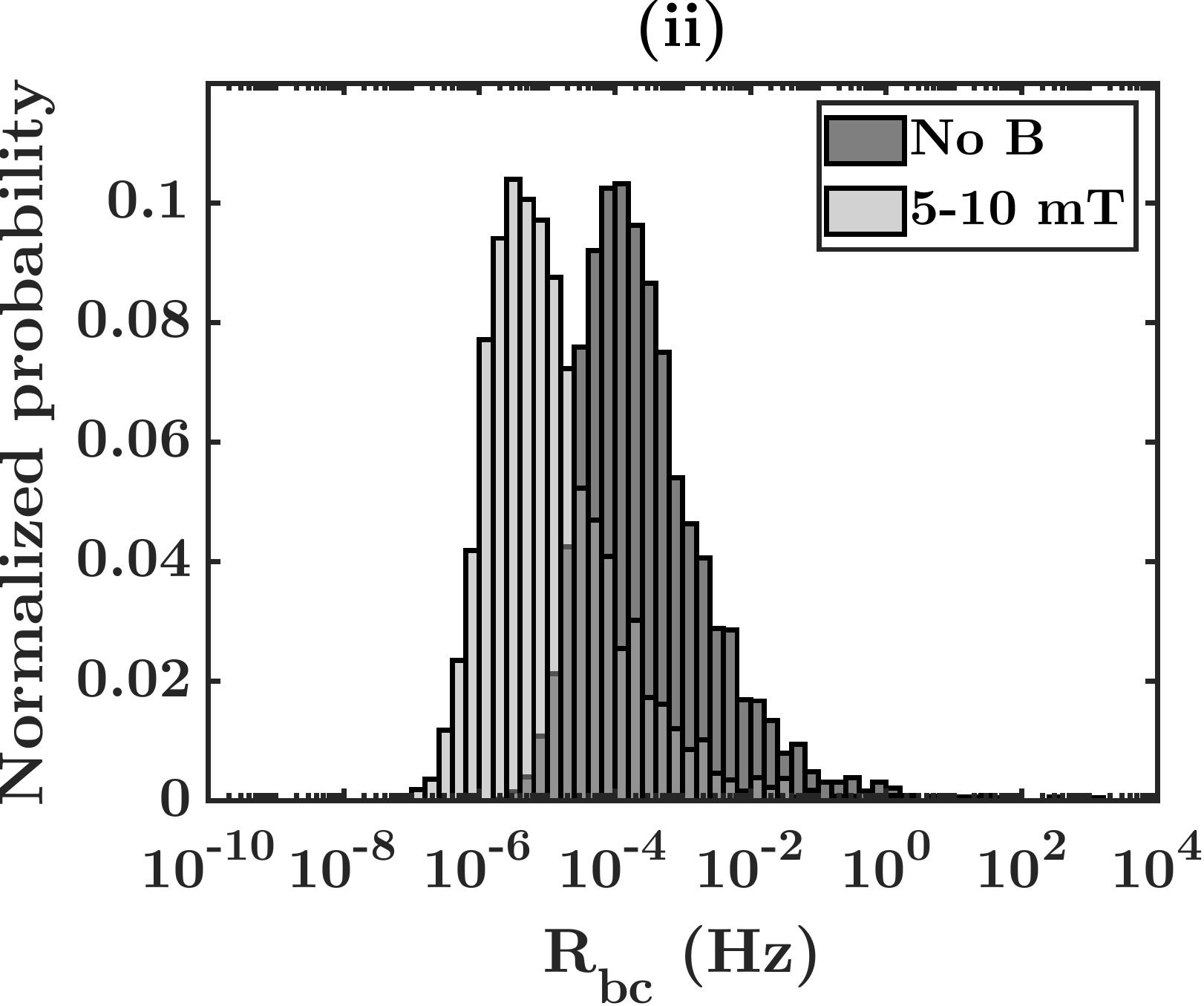}}
\vfill
\subfloat[]{\includegraphics[width=0.4\linewidth]{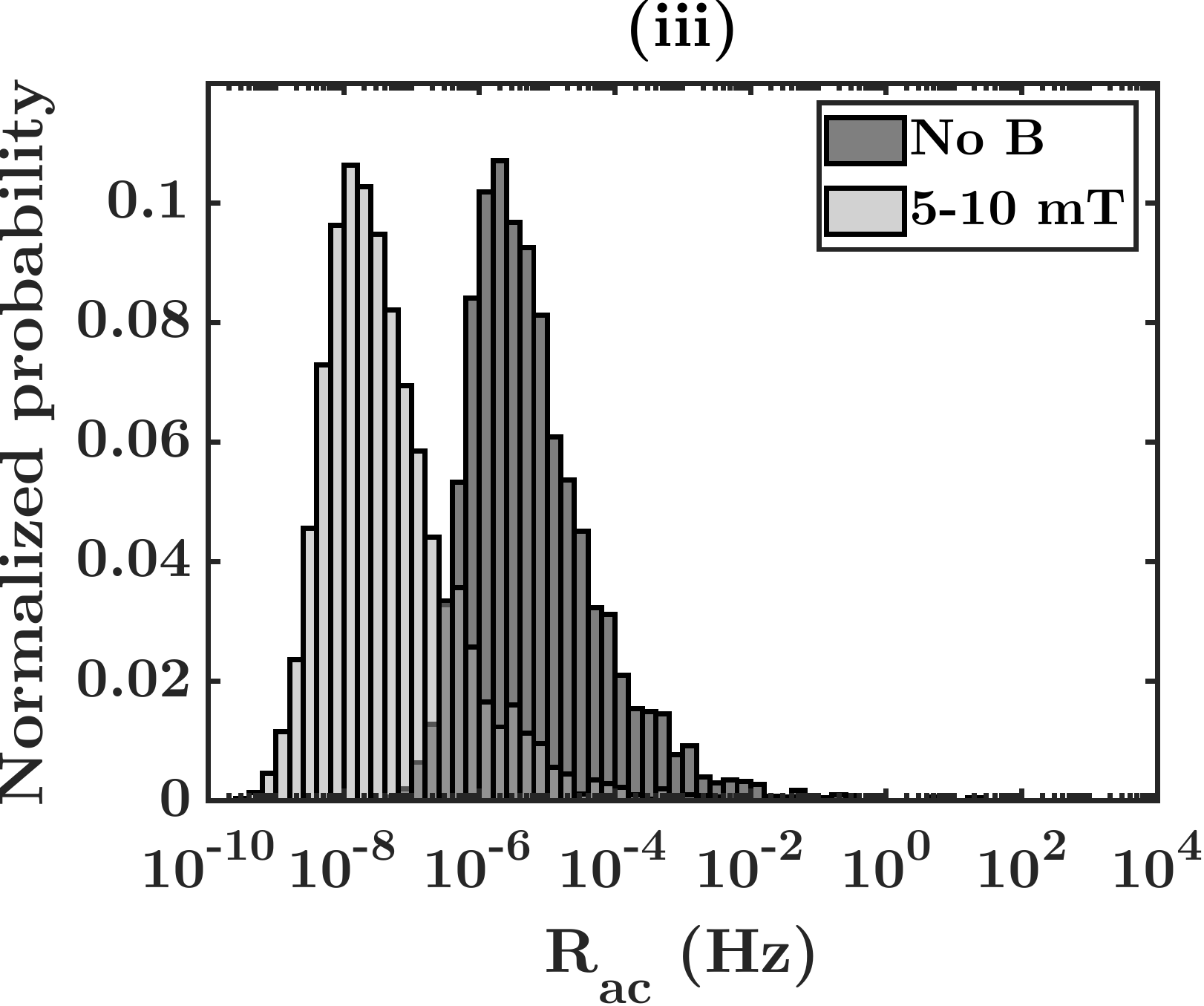}}
\hspace{1cm}
\subfloat[]{\includegraphics[width=0.4\linewidth]{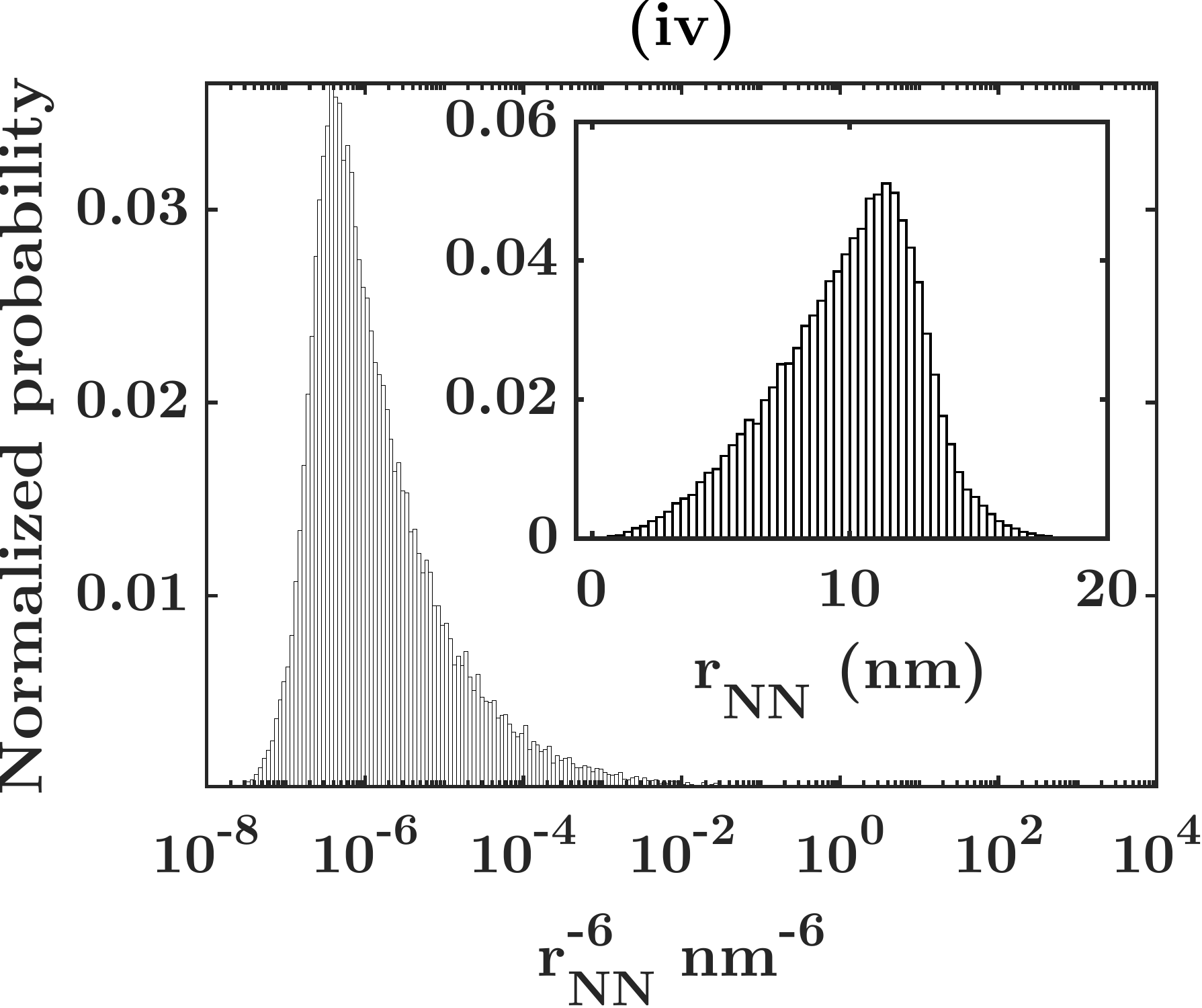}}
  \caption{Effect of magnetic field on the flip-flop rates. (i)-(iii) show a histogram of \(R_{ab}\), \(R_{bc}\) and \(R_{ac}\) respectively, for two cases : residual field \(<\)0.2 mT and external applied field of 5-10 mT. In the presence of residual field, \(R_{ab}\) peaks at  \(10^{-2}\) Hz while \(R_{bc}\) and \(R_{ac}\) peak at \(10^{-4}\) Hz and 2 x \(10^{-6}\) Hz. All the rates slow down by about two orders of magnitude with a field of 5-10 mT. The distribution of \(r^{-6}\)  for twenty closest neighbouring ions considered in the simulations is shown in (iv). The inset shows the histogram of \(r\), the ion-ion distance.}
  \label{fig:histogram_R}
\end{figure*}

\section{Results and Discussion} \label{sec:results_and_discussion} 
Results of population decay are shown in Figure \ref{fig:DecayFits}. We will first describe the relaxation dynamics in Figure \ref{fig:DecayFits}(i-ii), in the absence of an applied field. The figures show the experimental data with error bars indicating the weighted standard deviation of three data sets taken for decay after initializing the populations in each hyperfine level \(\pm \ket{a}\) (blue), \(\pm \ket{b}\) (red) and \(\pm \ket{c}\) (purple). There was no external magnetic field applied but previous measurements indicate that there is a residual field \(<\)0.2mT in our cryostat. Population decay for the first 5 ms is negligible so a moving average is performed up to this point and subsequently, the population is normalized with respect to this point. Decay from \(\pm \ket{c}\) is slower than \(\pm \ket{a}\) or \(\pm \ket{b}\) so one can expect \(R_{ac}\) and \(R_{bc}\) to be lower than \(R_{ab}\). In other words, ions in \(\pm \ket{c}\) flip-flop with those in either of the other levels at a much slower rate. As described earlier in Section \ref{sec:SpinRelaxation}, a single average value for each of \(R_{ab}, R_{bc}, R_{ac}\) is typically used to describe the relaxation of all ions in the crystal. Thus, all ions relax bi-exponentially with \( N=1\) in Equations (\ref{NaClassI}),(\ref{NbClassV}),(\ref{NcClassIX}) in Appendix \ref{sec:rateeqns}.  As an example of this ´macroscopic' model, we attempted a bi-exponential fit to our data, plotted using dashed-dotted lines in Figure \ref{fig:DecayFits}(i). The best fit obtained for \(\pm \ket{a}\), \(\pm \ket{b}\), \(\pm \ket{c}\) respectively was \(0.52e^{-t/5.52}+0.48e^{-t/2193}\) (blue), \(0.48e^{-t/5.52}+0.52e^{-t/2193}\) (red) and \(0.03e^{-t/5.52}+0.97e^{-t/2193}\) (purple), where time \(t\) is in seconds. While these curves fit well to many data points, several data points do not follow the fits especially \(\pm \ket{a}\) and \(\pm \ket{b}\). 

Each decay curve obtained in Figure \ref{fig:DecayFits} is in fact, an average of many exponential decays of different ions within the 1 MHz peak shown in Figure \ref{fig:energylevels}(ii). Each ion may have a different flip-flop rate for a given transition, depending on its position and orientation in the crystal.  In the microscopic model, the effective decay is instead an average of the bi-exponential decay of many ions in the crystal, shown as the solid colored (blue, red, purple) lines in Figure \ref{fig:DecayFits}(ii). These are the simulations which evaluate population according to steps detailed in Section \ref{sec:dipole_dipole} and they match the experimental data quite well. The solid colored lines in Figure  \ref{fig:DecayFits}(ii) and (iii) show the fits from simulation of our microscopic model in the absence and presence of magnetic field respectively. In addition to the list of steps in simulations described in Section \ref{sec:dipole_dipole}, a few more steps were followed in order to be able to compare the simulations with the experiments:
    \begin{enumerate}
     \item Using the rates \(R_{ab},R_{bc},R_{ac}\), population decay in levels \(\pm\ket{a}\),\(\pm\ket{b}\),\(\pm\ket{c}\) is calculated using respectively, Equations (\ref{NaClassI}),(\ref{NbClassV}),(\ref{NcClassIX}) derived in Appendix \ref{sec:rateeqns}.
    \item Steps [1-3] from the list in Section \ref{sec:dipole_dipole} are repeated for \(i=2,3...N\), where \(N\) is the number of ions in the sphere. An average decay of `\(N\)' ions gives a single decay curve describing the decay of all ions in the crystal. These are the solid colored lines in Figure \ref{fig:DecayFits}(iii).
    \item All of the above steps are repeated for data with an external magnetic field.
    \item Parameters \(\Gamma_{ab}, \Gamma_{bc},\Gamma_{ac}, \kappa_{ab}, \kappa_{bc}\) and \(\kappa_{ac}\) are optimized to match the experimental data. 
\end{enumerate}

 The experiments show little difference between the decay from \(\pm \ket{a}\) and \(\pm \ket{b}\), indicating that ions occupying these states have the strongest magnetic dipole-dipole interaction. This is shown by the blue and red solid lines almost overlapping with each other in \ref{fig:DecayFits}(i). The optimized values of spin inhomogeneous linewidths, \(\Gamma_{ab}, \Gamma_{bc}\) and \(\Gamma_{ac}\) were found to be  0.618, 3.309 and 2.664 kHz respectively. We choose ions in a sphere of radius 100 nm for the simulations, thus the fitted linewidths represent the local spin inhomogeneity and can be less than the measured values  \(\Gamma_{ab} =\) 50.5 kHz, \(\Gamma_{bc} = \)75.4 kHz \cite{lovric_spin_2012} in a bulk crystal. The optimization is fairly insensitive to \(\Gamma_{ac}\) and the relaxation is predominantly governed by the rates \(R_{ab}\) and \(R_{bc}\). We now try to understand why the fitted values of \(\Gamma_{ab}\) and \(\Gamma_{bc}\) differ by a factor of $\sim$4.4. A possible contribution to spin inhomogeneity is inhomogeneity in the g-tensor which stem from strains or defects \cite{RevModPhys.41.82}. Local inhomogeneity in spin could also be due to magnetic dipole-dipole interactions between a Pr ion with its neighbouring Pr ions of the type given by Equation \ref{dipole dipole hamil}. If the hyperfine wavefunctions \(\pm \ket{a}, \pm \ket{b}\) and \(\pm \ket{c}\) were composed of pure \(\pm \ket{\frac{1}{2}g}, \pm \ket{\frac{3}{2}g}\) and \(\pm \ket{\frac{5}{2}g}\) states, then the shift in hyperfine frequencies due to interaction between a pair of Pr ions scales linearly with the quantum number \(m_I = \frac{1}{2}, \frac{3}{2}, \frac{5}{2}\). Thus, the frequency shift of one ion in a pair occupying \(\pm \ket{\frac{5}{2}g}\) and \(\pm \ket{\frac{3}{2}g}\) is five times larger than that of another ion in a pair occupying \(\pm \ket{\frac{1}{2}g}\) and \(\pm \ket{\frac{3}{2}g}\). Furthermore, the effect of external magnetic field is largest on \(\pm\ket{c}\) since it undergoes a larger Zeeman shift compared to \(\pm \ket{a}\), as seen in Figure \ref{fig:spectrawithwithoutB} (ii) and (vi) in Appendix \ref{sec:spectrawithwithoutB}. This could explain why \(\Gamma_{bc}\) is $\sim$4.4 x  \(\Gamma_{ab}\) even though \(\pm \ket{a}, \pm \ket{b}\) and \(\pm \ket{c}\) are actually an admixture of the pure hyperfine states.

For the experiments with magnetic field in Figure \ref{fig:DecayFits}(iii), the field is put on after the population initialization step and it takes a few seconds for the field to ramp up to the set value. Thus, the simulation evolves the population until the dotted line at 4.6 seconds assuming there is no external field and normalizes the data so that the population at the time corresponding to the dotted line in (iii) equals the population in (ii). Some of the data points in \(\pm \ket{c}\) in (iii) before the dotted line show population greater than 1. This is an experimental artefact and occurs because the peak corresponding to these points, `3' in Figure [\ref{fig:energylevels}] split in the presence of field due to nuclear Zeeman effect and thus a different spectral region is chosen for evaluating population before and after the peak has split. This is shown in detail in Appendix [\ref{sec:spectrawithwithoutB}].%

After the dotted line, it is assumed that the magnetic field has reached the set value and the simulation evolves the population by including the phenomenological terms, \(\kappa_{ab},\kappa_{bc}\) and \(\kappa_{ac}\) introduced earlier in Fermi's rule [\ref{Fermi rule}]. The optimized values were found to be respectively, 2.6, 3.6 and 1.5 with a field between 5-10 mT. Figure \ref{fig:histogram_R} shows the effect of magnetic field on the calculated rates, where (i),(ii) and (iii) show the histogram of \(R_{ab}\), \(R_{bc}\) and \(R_{ac}\) respectively with and without a magnetic field. \(R_{ab}\) (with no applied field) is spread over a distribution ranging from \(10^{-4} - 10^{2}\) Hz and peaks at \(10^{-2}\) Hz. \(R_{bc}\) is slower, ranging from \(10^{-6} - 1\) Hz and peaks at \(10^{-4}\) Hz, with no field while \(R_{ac}\) is slowest, ranging from \(10^{-7} - 1\) Hz and peaks at 2 x \(10^{-6}\) Hz. All three rates slow down by two orders of magnitude with a field of 5 - 10 mT. The distribution of rates shown in (i) - (iii) follow from the distribution of \(r^{-6}_{NN}\) shown in (iv) and the inset shows the distribution of \(r_{NN}\), where \(r\) is the distance to any of the twenty closest neighbours of any ion considered in the simulations.

To understand why the rates slow down in a magnetic field, one can infer from Equation (\ref{Fermi rule}) that the cause could either be evolution of matrix elements in the dipole-dipole interaction term or a change in density of states \(f(E)\). While the matrix elements do not change appreciably with a small field of 5 - 10 mT, the density of states changes drastically due to the decrease in homogeneous linewidth by more than a factor of ten, as measured in Ref.\cite{FRAVAL2004347} and this is attributed to minimizing spin flips of the neighbouring Y ions. In the absence of an external field, the magnetic field experienced by the core Y ions is due to the local Pr ion, which is of the order of \( \sim 0.1\) mT and, a change in the spin state of Pr flips the spin state of Y ions. Thus, dephasing of Pr ions is dominated by neighbouring Y flips in the core. When the external field significantly exceeds the field due to the local Pr ion, such flips are minimized. Another factor contributing to the change in density of states is the increase in the spin inhomogeneous linewidth, characterized by the fitting parameters \(\kappa_{ab},\kappa_{bc}\) and \(\kappa_{ac}\). A linear increase in spin inhomogeneous linewidths has also been reported in  Nd\(^{3+}\):Y\(_2\)SiO\(_5\) \cite{cruzeiro_spectral_2017} and in erbium doped glass fibers \cite{PhysRevB.92.241111}. Measurements of spin linewidths as a function of magnetic field has partly been done in some Kramers ions \cite{WELINSKI201769} and similar measurements in Pr\(^{3+}\):Y\(_2\)SiO\(_5\) may shed more light on this explanation but such data is unavailable at this point. 
     
We conclude this section by noting that there are two conditions that need to be satisfied for two Pr ions to flip-flop : they need to be close to each other in the crystal and they also need to be spectrally close in the spin inhomogeneous profile. In our model, we take an average value for the spin inhomogeneity and model the ion-ion distance as a distribution. One could also model the spin inhomogeneity as a distribution, for example by including the effect of the local magnetic field around each Pr ion. The term \(\mathbf{B}\) in Equation (\ref{spin hamiltonian}) could be replaced by \(\mathbf{B_{total}=B_{ext} + B_{local}}\) so that each ion has a unique Spin Hamiltonian, resulting in a distribution of Zeeman frequencies of Pr ion. 

\section{Conclusion}
We have presented a method to model microscopic effects of flip-flop interactions between individual ions in a rare-earth-ion doped crystal. We have simulated a random doping based on the crystal structure of the host, where the position and orientation of all ions is known. Every dopant ion is situated in a unique position and orientation with respect to its neighbours so the ion-ion distance is a distribution and the flip-flop rate of any ion with its neighbours is different owing to this distribution. We apply this model to experiments of population decay of ground state hyperfine levels in Pr\(^{3+}\):Y\(_2\)SiO\(_5\). The experimental method used is an alternative to methods used in earlier works. The collective relaxation dynamics of all ions probed in the crystal is an average sum of many exponential decays of different ions. Thus, the flip-flop rate between two hyperfine levels is a distribution of rates rather than one average rate describing the dynamics of all ions. 

The fastest rate is \(R_{ab}\) between the levels \(\pm \ket{a}\) and \(\pm \ket{b}\), whose distribution has a peak at \(10^{-2}\) Hz while \(R_{bc}\) and \(R_{ac}\) have a peak at \(10^{-4}\) and 2 x \(10^{-6}\) Hz respectively, in the presence of a residual field \(<\)0.2 mT. All the rates decrease by 2 orders of magnitude upon applying an external field of 5 - 10 mT and the reason could be a combination of an order of magnitude decrease in the spin homogeneous linewidths and an increase in spin inhomogeneous linewidths\cite{cruzeiro_spectral_2017,car_optical_2019}. An improvement to the model could be to include the effect of differences in the local magnetic field around each dopant ion. Nonetheless, our model serves as a general tool to calculate other kinds of interactions at the microscopic level. It could be used to study the dynamics of other rare-earth ions in different materials as well.

\bibliography{SpinDynamics}

\begin{acknowledgments}
We thank Prof. Peter Samuelsson, Dr. Mikael Afzelius for useful discussions and Dr. Sebastian P. Horvath for providing partial code to calculate the spin Hamiltonian. This work was supported by Knut and Alice Wallenberg Foundation (KAW 2016.0081), Wallenberg Center for Quantum Technology (WACQT) funded by the Knut and Alice Wallenberg Foundation (KAW 2017.0449), Swedish Research Council (no. 2016-05121, no. 2019-04949) and European Union FETFLAG program, Grant No. 820391 (SQUARE).

\end{acknowledgments}

\clearpage

\section{Appendix}
\subsection{Density of states for transitions between levels in  two different continuum of states} \label{densityofstates}
In this section, we give details of how Fermi's rule is applied to the case of a flip-flop transition between two homogeneously broadened levels centered around different energies. We first start with Fermi's rule for a transition between two discrete levels and then extend it to the case of transition between a level in a continuum of states with finite width to another continuum of states. Finally, we apply this to flip-flop transitions, where each state is a two-level system.

Let us start with two levels \(\ket{1}\) and \(\ket{2}\) with energies \(E_1\) and \(E_2\) respectively. The transition rate from a discrete state with energy \(E_1\) in \(\ket{1}\) to \(E_2\) in \(\ket{2}\) under a perturbation \(H'\) is given by Fermi's rule :

\begin{equation*}
    R_{E_1 \rightarrow E_2} = \frac{2\pi}{\hbar} |\mel{1}{H'}{2}|^2 \delta(E_1 - E_2)
\end{equation*}

Assuming the matrix element \(|\mel{1}{H'}{2}|\) is independent of the energies \(E_1\) and \(E_2\), the total transition rate from the continuum of states in \(\ket{1} \rightarrow \ket{2}\) is given by integrating  \(R_{E_1 \rightarrow E_2}\) over the density of states for both the initial and final energies :

\begin{align}
\begin{split}
     R_{\ket{1} \rightarrow \ket{2}} =& \int dE_{1 }dE_{2 } \rho_1(E_1) \rho_2(E_2)  R_{E_1 \rightarrow E_2}  \\
    =& \frac{2\pi}{\hbar} |\mel{1}{H'}{2}|^2 \int dE_{ } \rho_1(E) \rho_2(E) 
\end{split}
\end{align}

Further, assume that both \(\rho_1\) and \(\rho_2\) have a normalized Lorentzian lineshape centered around \(\epsilon_1\) and \(\epsilon_2\) with homogeneous HWHM (half width at half maxima) \(\Delta_1\) and \(\Delta_2\) respectively such that, \(\rho_l(E) = \frac{1}{\pi}\frac{\Delta_l}{\Delta_l^2 + (E-\epsilon_l)^2} \), where \(l = 1,2\). The rate is then :

\begin{equation} \label{fermi_ge}
     R_{\ket{1} \rightarrow \ket{2}} = \frac{2\pi}{\hbar} |\mel{1}{H'}{2}|^2 \biggl [\frac{1}{\pi} \frac{\Delta_1 + \Delta_2}{(\Delta_1 + \Delta_2)^2 + (\epsilon_1 - \epsilon_2)^2} \biggr ]
\end{equation}

For the case of flip-flop transitions, perturbation \(H'\) is the magnetic dipole-dipole Hamiltonian between two ions `\(i\)' and `\(j\)' : \(H_{dd}^{ij}\). \(\ket{1}\) is the two-level system \(\ket{x^i}\otimes\ket{y^j}\), which flips to  \(\ket{2}\)  i.e., \(\ket{y^i}\otimes\ket{x^j}\) due to  \(H_{dd}^{ij}\), where \(\ket{x},\ket{y}\) are the wavefunctions of the hyperfine levels \(+\ket{a},-\ket{a}...-\ket{c}\) (shown in Figure \ref{fig:energylevels}). In the equation \ref{fermi_ge} above, \(\epsilon_1\) and \(\epsilon_2\) are the energies of \(\ket{x^i}\otimes\ket{y^j}\) and \(\ket{y^i}\otimes\ket{x^i}\) respectively and \(\Delta_1\) and \(\Delta_2\) are their respective homogeneous HWHM of Lorentzian lineshapes. The lineshape of such a transition is given by the  convolution of the two Lorentzian lineshapes of the individual levels and is another Lorentzian with HWHM  \(\Delta_1 + \Delta_2\), centered at \(\epsilon_1 + \epsilon_2\). From Ref. \cite{PhysRevLett.92.077601,FRAVAL2004347}, we take \(2\Delta_1 = \frac{1}{\pi h T_2}\), where \(T_2\) is the spin coherence time at zero magnetic field and it is equal to 0.5 ms (6 ms with a field of 5-10 mT). Considering \(\Delta_1 = \Delta_2\),  HWHM is \(\Delta = \frac{1}{\pi h T_2}\), which is just the homogeneous linewidth of the transition \(\ket{x} \leftrightarrow \ket{y}\). Also, \((\epsilon_1 - \epsilon_2) = h \Gamma_{xy}\), where \(\Gamma_{xy}\) is the spin inhomogeneity .

Writing Equation \ref{fermi_ge} in terms of frequencies and replacing \(\Delta_1 + \Delta_2\) by \(h\Gamma_{hom}\) and \((\epsilon_1 - \epsilon_2)\) by  \(h\Gamma_{xy}\), we get the final expression :

\begin{equation} \label{fermi_ge_freq}
     R_{\ket{x} \rightarrow \ket{y}} = \frac{2\pi}{\hbar} |\mel{y^i\otimes x^j}{H_{dd}^{ij}}{x^i \otimes y^j}|^2 \biggl [\frac{1}{\pi h} \frac{\Gamma_{hom}}{\Gamma_{hom}^2 + \Gamma_{xy}^2} \biggr ]
\end{equation}

Thus we can use the last term on the right of the above equation as the form of density of states \(f(E)\). Similar expressions have been used to describe transitions between broadened states in different quantum wells, as described in Section 3.3 of Ref. \cite{wacker_semiconductor_2002}. In Equation \ref{Fermi rule} in the main text, the homogeneous linewidths are functions of magnetic field and we also have a phenomenological factor \(\kappa_{xy}\) to describe the increase in inhomogeneous linewidths in the presence of magnetic field. 

\subsection{Reduction of rates from 15 to 3 and optimization of parameters in simulations}
\label{reduction_of_rates}
In the absence of an external field, there are three ground state hyperfine levels in Pr\(^{3+}\):Y\(_2\)SiO\(_5\) and each splits into two in the presence of a field as shown in Figure \ref{fig:spinspin interaction}. There can be fifteen unique rates due to magnetic dipole transitions in this case. Flip-flop interactions where initial and final state are the same or only change parity, for example \(R_{+\ket{b} \leftrightarrow -\ket{b}}\) are ignored since we do not measure these individually in our experiments. Thus, the simulations calculate twelve unique rates. However, our experiments are designed to measure only three rates \(R_{\pm \ket{a} \leftrightarrow \pm \ket{b}}, R_{\pm \ket{b} \leftrightarrow \pm \ket{c}}, R_{\pm \ket{a} \leftrightarrow \pm \ket{c}}\), or  referred to in the main article as \(R_{ab}, R_{bc}, R_{ac}\). Each of the rates is divided by 6 since the neighbouring ion can only be in one of the six hyperfine levels. So we are left with the task of reducing the twelve rates from simulations down to three. 

\begin{figure*}
\centering
\includegraphics[width=0.9\linewidth]{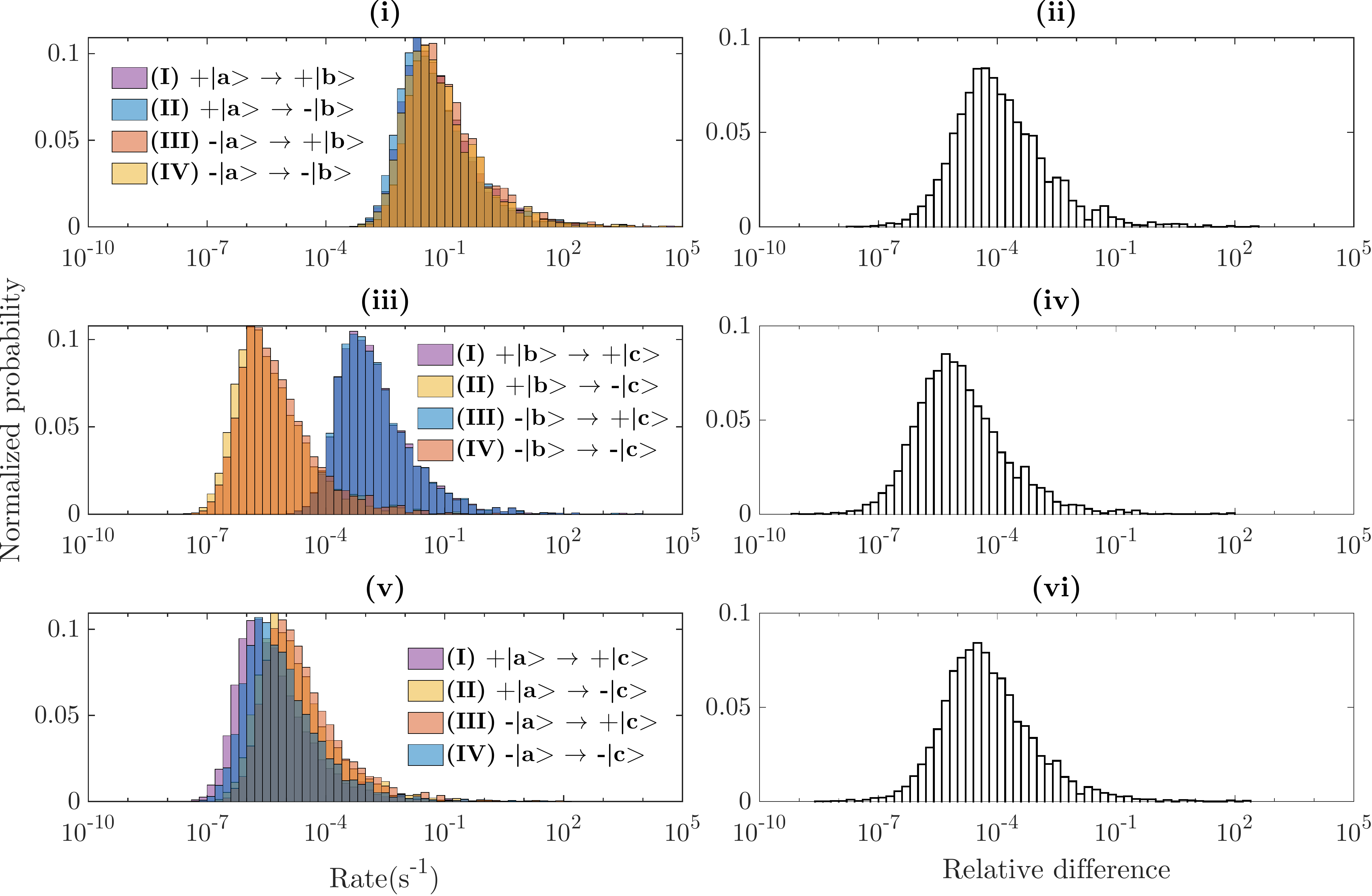}
\caption{Histogram of rates of all transitions calculated between each pair of levels without an external magnetic field : (i)\(\pm \ket{a},\pm \ket{b}\) (iii) \(\pm \ket{b},\pm \ket{c}\) and (v) \(\pm \ket{a},\pm \ket{c}\). Relative difference between the quantities (I+II), (III+IV) is shown in sub figures (ii),(iv),(vi). Average of (I+II) and (III+IV) is taken to give three effective rates \(R_{ab}, R_{bc}\) and \(R_{ac}.\)}
\label{fig:Allrates_splitup}
\end{figure*}

We divide the rates into three categories, each involving the pair of levels \(\pm \ket{a}\) and \(\pm \ket{b}\), \(\pm \ket{b}\) and \(\pm \ket{c},\)  \(\pm \ket{a}\) and \(\pm \ket{c}\).  Histogram of rates involving the transitions in each pair are shown in Figure \ref{fig:Allrates_splitup} (i),(iii),(v). We first consider (i).
\begin{itemize}
    \item Transitions originating from the same level are added together. For example, The rates \(+\ket{a} \rightarrow +\ket{b}\) and  \(+\ket{a} \rightarrow -\ket{b}\) (labelled as (I) and (II) in the Figure \ref{fig:Allrates_splitup} (i) ). Similarly (III) and (IV), \(-\ket{a} \rightarrow +\ket{b}\) and \(-\ket{a} \rightarrow -\ket{b}\) are added together.
    \end{itemize}
    
    \begin{itemize}
    \item Since we cannot distinguish \(+\ket{a} \rightarrow +\ket{b}\) from \(-\ket{a} \rightarrow +\ket{b}\), we take the average of (I+II) and (III+IV), which is reasonable since they are very similar and in experiments, the split peaks appear to decay at the same rate (see Figure \ref{fig:spectrawithwithoutB}(iv) and (vi) for the case of \(\pm\ket{b}\) and \(\pm\ket{c}\). This total rate is called as \(R_{ab}\). The relative difference between the two quantities, calculated as \([(\text{I+II)-(III+IV)}]/\text{Mean}[R_{ab}]\) is shown in Figure \ref{fig:Allrates_splitup}(ii). 
  \end{itemize}

A similar argument is applied to (iii-iv) and (v-vi) in Figure \ref{fig:Allrates_splitup}, thus the rates are reduced from twelve to three.

Optimization of the parameters \(\Gamma_{ab}, \Gamma_{bc}, \Gamma_{ac}, \kappa_{ab}, \kappa_{bc}\) and \(\kappa_{ac}\) was done using Global Search and fminsearch functions in Matlab. The cost function, which is a measure of deviation in population between simulated model (\(p_{m}\)) and experiments (\(p_{e}\)) is minimized to have the lowest 'score' simultaneously for decay in all three hyperfine levels as well as for all values of magnetic field :

\begin{equation*}
    score = \sum _{\substack{ \forall \textrm{ times} \\ \forall \pm\ket{a}, \\ \forall \mathbf{ B}} } (\frac{p_{e}- p_{m}}{w*p_{e}})^2 
\end{equation*}

 For data without external magnetic field, \(w\) is the standard deviation of three data sets taken at each time point and for cases with magnetic field, \(w\) is the standard deviation of three data sets, each taken at 5, 7 and 10 mT. In total, there are six parameters describing the decay : \(\Gamma_{ab}, \Gamma_{bc}, \Gamma_{ac}, \kappa_{ab},\kappa_{bc},\kappa_{ac}\). The score is most sensitive to value of \(\Gamma_{ab}\). A change of \(\pm 5 \%\) in \(\Gamma_{ab}\) changes the score by \(\sim 10 \%\) while the same change in \(\Gamma_{bc}\) or \(\kappa_{ab}\) changes the score by \(\sim 3 \%\). Changing \(\kappa_{bc}\) by \(\pm 5 \%\) results in a change in score by \(\sim 1.5\%\).  Thus, the relaxation is mostly governed by the fastest rate, \(\Gamma_{ab}\). It was also checked that increasing the size of the sphere or number of neighbours did not further improve the score appreciably, thus it is sufficient to calculate the effect of twenty nearest neighbours on each other and in a YSO crystal, this distance varies between 1-20 nm.

 \begin{table*}
\caption{Conditions of population after initializing ions in \(\frac{1}{2}\)g state at 0MHz}
\label{tab:table1}
\begin{ruledtabular}
\begin{tabular}{cccccccc}
 & &\multicolumn{3}{c}{\(f_{peak}=0\) MHz}&\multicolumn{3}{c}{\(f_{background}=2\)MHz}\\
 Class &Transition probed at 0 MHz&$n_a$&$n_b$&$n_c$&$n_a$
&$n_b$&$n_c$\\ \hline
 I&$\pm \ket{a}\rightarrow\pm \ket{a_e}$&1 &0 & 0& 0&0  &1 \\
 II&$\pm \ket{a}\rightarrow\pm \ket{b_e}$&0 &0 &1 & 0& 0 &1 \\
 III&$\pm \ket{a}\rightarrow\pm \ket{c_e}$& 0&0 &1 & 0& 0 &1 \\
 IV&$\pm \ket{b}\rightarrow\pm \ket{a_e}$&0 & 0& 1&0 & 0 &1 \\
V&$\pm \ket{b}\rightarrow\pm \ket{b_e}$&1 &0 &0 &1 & 0 &0 \\
 VI&$\pm \ket{b}\rightarrow\pm \ket{c_e}$&1 &0 & 0& 1& 0& 0\\
 VII&$\pm \ket{c}\rightarrow\pm \ket{a_e}$&1 &0 & 0& 1&0  &0 \\
 VIII&$\pm \ket{c}\rightarrow\pm \ket{b_e}$&1 &0 &0 & 1& 0 &0 \\
  IX&$\pm \ket{c}\rightarrow\pm \ket{c_e}$&1 &0 & 0& 1&0 & 0

\end{tabular}
\end{ruledtabular}
\end{table*}

\begin{figure*}
\centering
  \includegraphics[width = 0.8\linewidth]{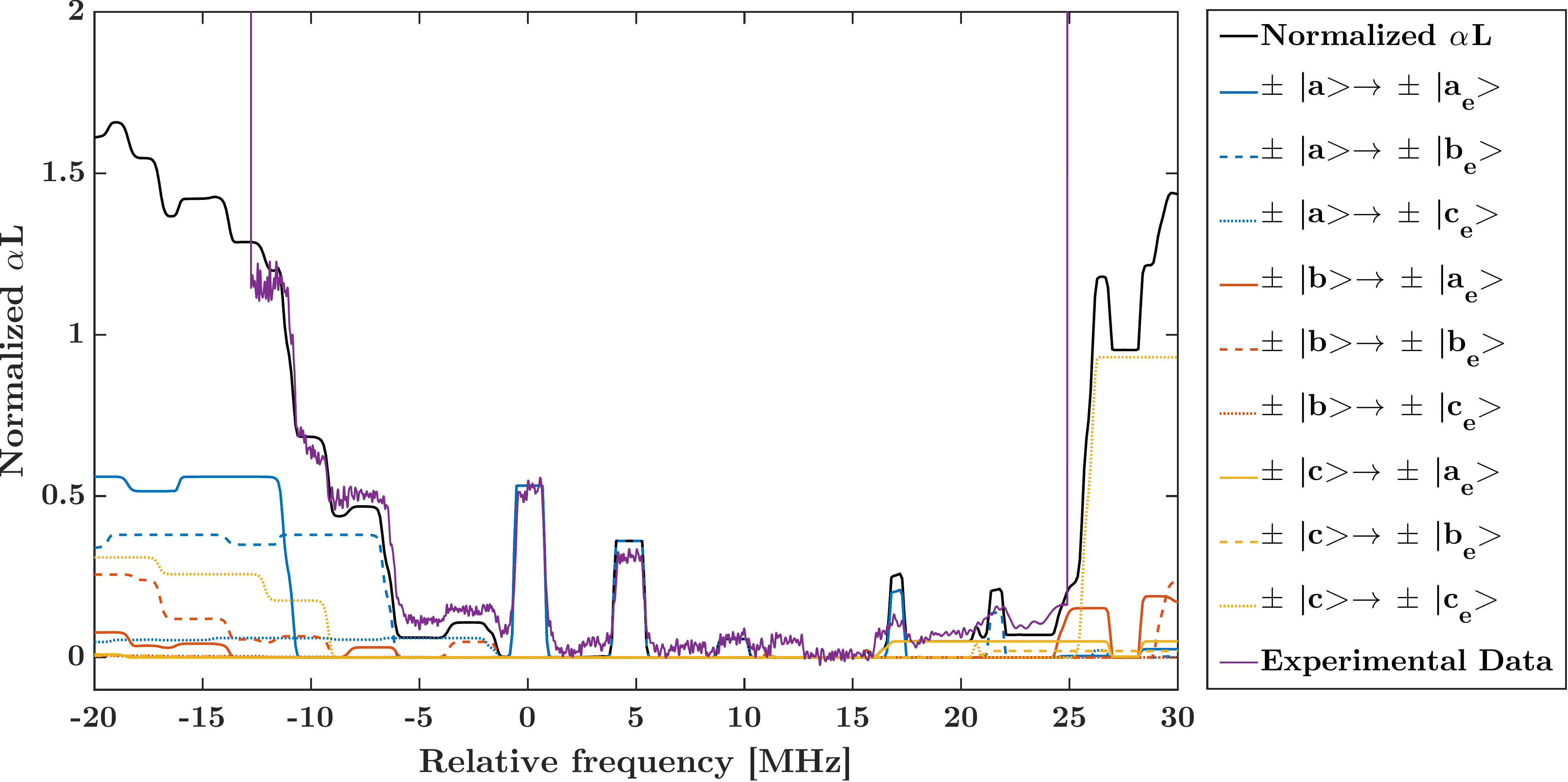}
  \caption{Simulation of absorption spectrum after initializing population in \(\pm \ket{a}\). The black line shows the total absorption while colored lines show the absorption corresponding to a particular transition, indicated in the legend. Note the three peaks at 0, 4.6, 9.4 MHz which have ions absorbing only from \(\pm \ket{a}\), indicating successful isolation of one class from the other eight classes. Purple line corresponds to the first readout at 1ms after the initialization step in the experiments.}
  \label{fig:simulation_absorption}
\end{figure*}

\subsection{Rate Equations and Initial Conditions of population} \label{sec:rateeqns}
In this section, we derive the rate equations for a three level system \(\pm \ket{a},\pm \ket{b},\pm \ket{c}\) shown in Figure \ref{fig:energylevels} and elucidate the initial conditions of population for nine classes of ions. If \(N_a,N_b,N_c\) are populations normalized to the total population \(N\) and \(R_{ab}, R_{bc},R_{ac}\) are the flip-flop rates between the three hyperfine levels, the rate equations can be written as: 

\[\frac{dN_a}{dt}=R_{ab}N_b + R_{ac}N_c-(R_{ab}+R_{ac})N_a\]
\[\frac{dN_b}{dt}=R_{ab}N_a + R_{bc}N_c-(R_{ab}+R_{bc})N_b\]
\[\frac{dN_c}{dt}=R_{ac}N_a + R_{bc}N_b-(R_{ac}+R_{bc})N_c\]

  The solutions for arbitrary initial conditions \(N_a=n_a\), \(N_b=n_b\), \(N_c=n_c\) are :

\begin{align}
\begin{split} \label{Na eq}
&N_a(t)=  \frac{n_a + n_b + n_c}{3} + \frac{1}{6\sigma} \biggl\{e^{-t(K+\sigma)} \bigl[n_a A_1 \\
& +n_b A_3+n_c A_2+\sigma(2n_a-n_b-n_c)\bigr] -e^{-t(K-\sigma)}  \\ 
& \bigl[n_a A_1+n_b A_3+n_c A_2 -\sigma(2n_a-n_b-n_c) \bigl]  \biggr\}\\     
\end{split}
    \end{align}
    \begin{align}
\begin{split} \label{Nb eq}
  & N_b(t)= \frac{n_a + n_b + n_c}{3} + \frac{1}{6\sigma} \biggl\{e^{-t(K+\sigma)} \bigl[n_a A_3 \\
   & +n_b A_2+n_c A_1 +\sigma(2n_b-n_a-n_c)\bigr] -e^{-t(K-\sigma)}\\
   &\bigl[(n_a A_3+n_b A_2+n_c A_1-\sigma(2n_b -n_a-n_c) \bigr] \big\} \\ 
  \end{split}
\end{align}
\begin{align}
\begin{split} \label{Nc eq}
    &N_c(t)= \frac{n_a + n_b + n_c}{3} +  \frac{1}{6\sigma} \biggl\{e^{-t(K+\sigma)} \bigl[n_a A_2\\
    &+n_b A_1+n_c A_3 + \sigma(2n_c-n_b-n_a)\bigr] -e^{-t(K-\sigma)} \\
    &\bigl[n_a A_2+n_b A_1+n_c A_3 -\sigma(2n_c-n_b-n_a)\bigr] \biggr\}  \\ 
    \end{split}
\end{align}

where
\begin{align*}
A_1=& R_{ab}+R_{ac}-2R_{bc} \\
A_2=& R_{ab}+R_{bc}-2R_{ac} \\
A_3=& R_{ac}+R_{bc}-2R_{ab} \\
K=& R_{ab}+R_{ac}+R_{bc} \\
\begin{split}
\sigma=& \bigl\{ R_{ab}^2 +R_{ac}^2+R_{bc}^2 \\
 & -R_{ab}R_{ac} -R_{ab}R_{bc}-R_{ac}R_{bc} \bigr\}^{1/2}
    \end{split}\\
\end{align*}

Due to the optical inhomogenous broadening, the laser can couple to nine different transitions from the ground to excited state (shown in Figure \ref{fig:energylevels}) at a given frequency. To extract the individual rates \(R_{ab}, R_{bc},R_{ac}\) from a simple hole burning spectra, one would need to follow the evolution of all classes \cite{klieber_all-optical_2003} by summing over contributions from nine transitions for each of the three levels with three unknown initial conditions, thereby giving 30 unknowns. We simplify this by creating a transmission window using optical pumping and isolating peaks of ions in one hyperfine level within this window. A simulation in a six level system (with three ground state and three excited states) was performed to predict an absorption spectra after the initialization step. 

An example of initializing in \(\pm \ket{a}\), with a peak at 0 MHz corresponding to \(\pm \ket{a} \rightarrow \pm \ket{a_e}\) is shown in Figure \ref{fig:simulation_absorption}. It can be seen that the experimental data in purple matches quite well with the black line showing the simulation, indicating successful isolation of one class of ions from the other eight classes. We call this isolated group of ions as Class I, for which the starting conditions after the initialization are \(n_a = 1, n_b = 0, n_c = 0\) and the corresponding spectral background region at 2MHz has all Class I ions shelved in \(n_c\). The initial population conditions of the other eight classes for both the peak and background are charted out in Table [\ref{tab:table1}]. The conditions for all classes but one, 'I' are the same for both peak and background. Thus, we can subtract background from the peak for this class only using Equation (\ref{Na eq}) to calculate population decay in \( \pm \ket{a}\). This describes the decay of an ion `\(i\)'. To account for all ions \(i=1,2...,N\) in the sphere considered in the simulations, we take the average as follows:

\begin{align}  \label{NaClassI}
N_a(t) \bigg |_{ Class  I} &=  \frac{1}{N} \sum_{i}^{N} (N_{a,peak} - N_{a,background}) \nonumber\\
 &=  \frac{1}{N} \sum_{i}^{N}   \frac{1}{2\sigma^i} \left [ (\sigma^i + R_{ac}^i - R_{bc}^i)e^{-t(K^i+\sigma^i)}  \right. \nonumber \\ & + \left. (\sigma^i - R_{ac}^i + R_{bc}^i)e^{-t(K^i-\sigma^i)}\right] 
\end{align}

Similar considerations for peaks corresponding to \(\pm \ket{b} \rightarrow \pm \ket{b_e}\) at 14.8 MHz and \(\pm \ket{c} \rightarrow \pm \ket{c_e}\) at 36.9 MHz (peaks '2' and '3' respectively in Figure \ref{fig:energylevels} are given in Tables \ref{tab:table2} and \ref{tab:table3}. Coupled with simulations similar to Figure \ref{fig:simulation_absorption}, isolation of one class was ensured. The evolution of population \(N_b(t)\) and \(N_c(t)\) can also be obtained:
\begin{align}  \label{NbClassV}
N_b(t) \bigg |_{ Class  V} &=  \frac{1}{N} \sum_{i}^{N} (N_{b,peak} - N_{b,background}) \nonumber \\
 &=  \frac{1}{N} \sum_{i}^{N}  \frac{1}{2\sigma^i} \left [ (\sigma^i - R_{ac}^i + R_{bc}^i)e^{-t(K^i+\sigma^i)}  \right. \nonumber \\ & +\left. (\sigma^i + R_{ac}^i - R_{bc}^i)e^{-t(K^i-\sigma^i)}\right]  
\end{align}
\begin{align}  \label{NcClassIX}
N_c(t) \bigg |_{ Class  IX} &=  \frac{1}{N} \sum_{i}^{N} (N_{c,peak} - N_{c,background}) \nonumber \\
 &=  \frac{1}{N} \sum_{i}^{N}  \frac{1}{2\sigma^i} \left [ (\sigma^i + R_{ac}^i - R_{ab}^i)e^{-t(K^i+\sigma^i})  \right.\nonumber  \\ & +\left. (\sigma^i + R_{ab}^i - R_{ac}^i)e^{-t(K^i-\sigma^i)}\right] 
\end{align}

\begin{table*}
\caption{Conditions of population for initializing ions in \(\pm \ket{b}\) state at 14.7MHz}
\label{tab:table2}
\begin{ruledtabular}
\begin{tabular}{cccccccc}
 & &\multicolumn{3}{c}{\(f_{peak}=\)14.7 MHz}&\multicolumn{3}{c}{\(f_{background}=12.2\) MHz}\\
 Class &Transition probed at 14.7 MHz&$n_a$&$n_b$&$n_c$&$n_a$
&$n_b$&$n_c$\\ \hline
 I&$\pm \ket{a}\rightarrow\pm \ket{a_e}$&0 &0 &1 &0 &0  &1 \\
 II&$\pm \ket{a}\rightarrow\pm \ket{b_e}$&0 &0 &1 &0 & 0 &1 \\
 III&$\pm \ket{a}\rightarrow \pm \ket{c_e}$&0 & 0&1 & 0&0  &1 \\
 IV&$\pm \ket{b}\rightarrow\pm \ket{a_e}$&0 & 0& 1& 0& 0 &1 \\
 V&$\pm \ket{b}\rightarrow \pm \ket{b_e}$&0 &1 & 0& 0&0  &1 \\
 VI&$\pm \ket{b}\rightarrow \pm \ket{c_e}$&0 &0 &1 & 0& 0& 1\\
 VII&$\pm \ket{c}\rightarrow\pm \ket{a_e}$&1 & 0& 0& 1& 0 & 0\\
 VIII&$\pm \ket{c}\rightarrow \pm \ket{b_e}$&1 &0 & 0& 1&0 & 0\\
  IX&$\pm \ket{c}\rightarrow \pm \ket{c_e}$&1 &0 & 0& 1&0 &0 

\end{tabular}
\end{ruledtabular}
\end{table*}

\begin{table*}
\caption{Conditions of population for initializing ions in \(\pm \ket{c}\) state at 36.9 MHz}
\label{tab:table3}
\begin{ruledtabular}
\begin{tabular}{cccccccc}
 & &\multicolumn{3}{c}{\(f_{peak}=36.9\) MHz}&\multicolumn{3}{c}{\(f_{background}=38.9\) MHz}\\
 Class &Transition &$n_a$&$n_b$&$n_c$&$n_a$
&$n_b$&$n_c$\\ \hline
 I&$\pm \ket{a}\rightarrow\pm \ket{a_e}$&0 &0 &1 & 0& 0 &1 \\
 II&$\pm \ket{a}\rightarrow\pm \ket{b_e}$&0 &0 &1 &0 & 0 &1 \\
 III&$\pm \ket{a}\rightarrow\pm \ket{c_e}$& 0&0 &1 &0 &0  &1 \\
 IV&$\pm \ket{b}\rightarrow\pm \ket{a_e}$& 0&0 & 1&0 &0  &1 \\
 V&$\pm \ket{b}\rightarrow \pm \ket{b_e}$& 0&0 &1 &0 & 0 &1 \\
 VI&$\pm \ket{b}\rightarrow \pm \ket{c_e}$&0 &0 &1 &0 & 0& 1\\
 VII&$\pm \ket{c}\rightarrow\pm \ket{a_e}$&1 &0 &0 & 1& 0 & 0\\
 VIII&$\pm \ket{c}\rightarrow \pm \ket{b_e}$&1 &0 &0 & 1&0 & 0\\
  IX&$\pm \ket{c}\rightarrow \pm \ket{c_e}$& 0& 0&1 & 1& 0&0 
\end{tabular}
\end{ruledtabular}
\end{table*}

\subsection{Experimental Details} \label{sec:experimental_details}
The light source is a Coherent 699-21 dye laser, optically pumped at 532 nm by a Verdi-V6 solid state laser. The dye solution was made using Rhodamine 6G mixed with ethylene glycol and pumped at 4.2 bar and cooled to 10\degree C. All the pulses were shaped using an arbitrary waveform generator (Tektronix AWG520) and two AOMs: `AOM 1' AA.ST.200/B100/A0.5-vis in double pass (in the same configuration as described in \cite{rippe_experimental_2005}) and `AOM 2' A.ST.360/B200/A0.5-vis in single pass. Diffracted light from `AOM 2' is coupled to a polarization-maintaining fiber to another table with the cryostat. A beam sampler (90:10) is used to reflect some light onto a reference detector PD1 while transmitting majority of the light towards the crystal. Both detectors used were Thorlabs PDB150A. The absorption as a function of frequency is obtained by scanning over the desired frequency range with a rate of 1 MHz/\(\micro\)s with a weak probe. Due to the fast readout scan-rate, the transmission signal contains Free Induction Decay from each peak that needs to be deconvoluted as described in \cite{chang_recovery_2005}.

\subsection{Absorption spectra with and without magnetic field}\label{sec:spectrawithwithoutB}

\begin{figure}
\centering
\includegraphics[width=1\linewidth]{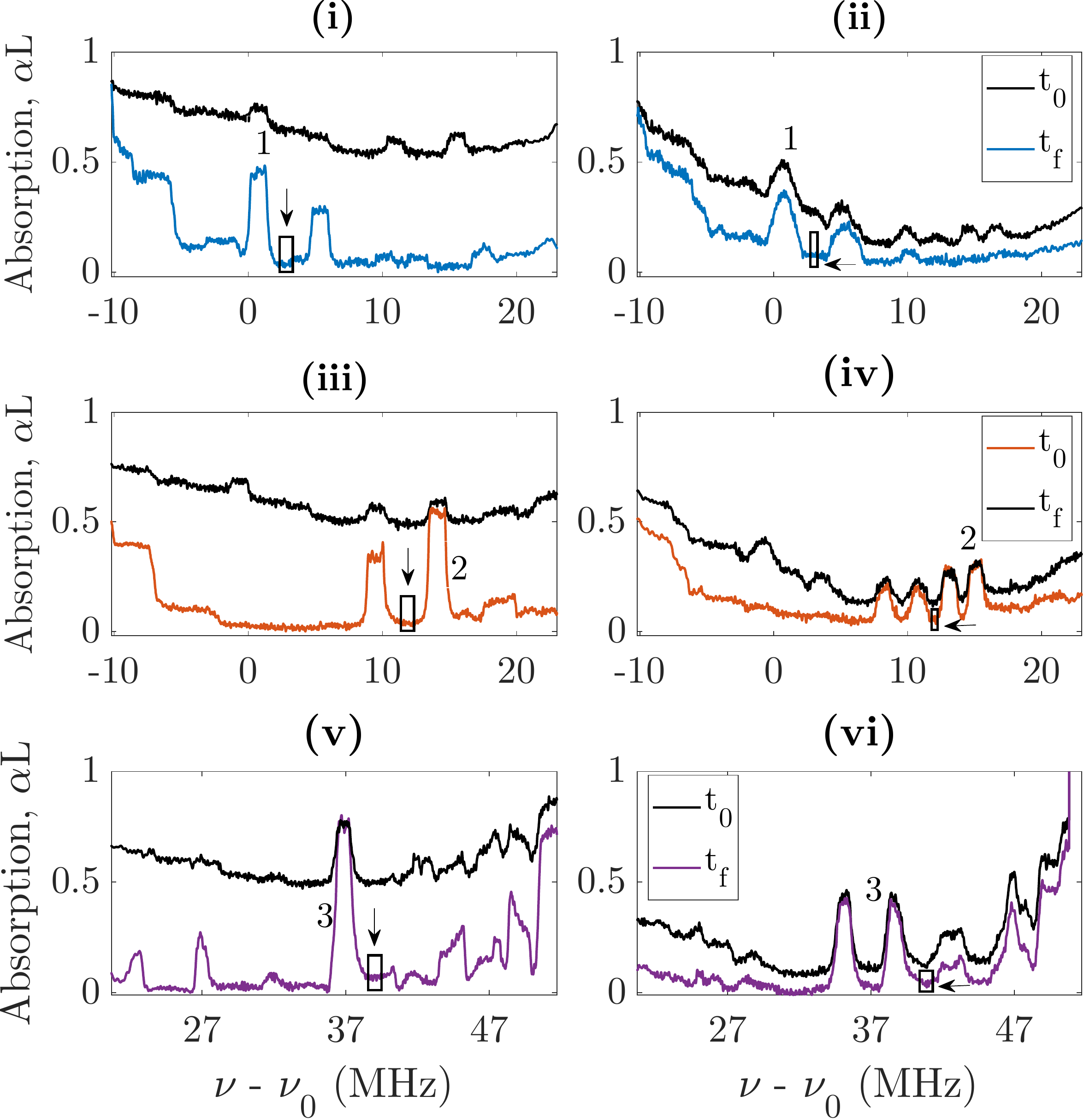}
\caption{Absorption structure after the first and last readout at times \(t_0\) and \(t_f\) respectively. (i),(iii),(v) : Population initialized at \(t_0\) in one of the three ground states shown in three colors : \(\pm \ket{a}\) in blue, \(\pm \ket{b}\) in red and \(\pm \ket{c}\) in purple. Peaks used for evaluating population and corresponding background region are indicated similar to Figure (\ref{fig:energylevels}). The black traces show the same absorption after the last readout at \(t_f\). The absorption increases over time as the ions which were optically pumped to other frequencies during initialization start to flip-flop into the transmission window. (ii),(iv),(vi) show the same experiments as (i),(iii),(v) but with an external field of 10 mT. }
 \label{fig:spectrawithwithoutB}
\end{figure}

Absorption spectra from the first and last readout are shown in Figure \ref{fig:spectrawithwithoutB}. (i),(iii) and (v) show the absorption after initializing the populations in \(\pm \ket{a}\) (blue), \(\pm \ket{b}\) (orange) and \(\pm \ket{c}\) (purple) respectively. All black traces show the absorption after the last readout at \(t_f\) = 2700s. Although no external magnetic field was applied in these cases, we expect there to be a stray field less than 0.2mT. Results of similar experiments with an external field of 10mT are shown in (ii),(iv),(vi) . The first readout \(t_0\) is at 3.8s for (ii) and (iv), 5.5s for (vi), which is also indicated by the vertical dashed line in Figure \ref{fig:DecayFits}. Peaks `2' and `3' in (iv) and (vi)  respectively split due to nuclear Zeeman effect. Although the absorption level of split peaks is roughly halved, the absorption level of their backgrounds are not the same as their counterparts in (iii) and (v). Due to this, some of the data points before the dotted line in Figure \ref{fig:DecayFits}(iii) are greater than 1. A different choice of background could perhaps have been better but the important information for simulations is what happens to the population after the magnetic field has reached the set value at the dotted line.

\end{document}